\documentclass[useAMS,usenatbib]{mn2e}
\usepackage{graphicx}
\usepackage{amssymb, amsmath}
\usepackage{times}
\usepackage{color}
\usepackage{lscape}
\usepackage[section]{placeins}
\usepackage{epstopdf}
\usepackage{epsf}

\newcommand{\wsp}{{\textcolor{white}0}}
\newcommand{\cchange}[1]{#1}

\begin{document}
\setlength{\parskip}{0pt}

\title[Small-scale structure]{Small-scale structure and the Lyman-$\alpha$ forest baryon acoustic oscillation feature}

\author[Hirata]{Christopher M. Hirata$^1$\\
$^1$Center for Cosmology and Astroparticle Physics (CCAPP), The Ohio State University, 191 West Woodruff Lane, Columbus, Ohio 43210, USA}

\date{28 October 2017} 
\pagerange{\pageref{firstpage}--\pageref{lastpage}} \pubyear{2017}
\maketitle
\label{firstpage}

\begin{abstract}
The baryon-acoustic oscillation (BAO) feature in the Lyman-$\alpha$ forest is a key probe of the cosmic expansion rate at redshifts $z\sim 2.5$, well before dark energy is believed to have become significant. A key advantage of the BAO as a standard ruler is that it is a sharp feature and hence is more robust against broadband systematic effects than other cosmological probes. However, if the Lyman-$\alpha$ forest transmission is sensitive to the initial streaming velocity of the baryons relative to the dark matter, then the BAO peak position can be shifted. Here we investigate this sensitivity using a suite of hydrodynamic simulations of small regions of the intergalactic medium with a range of box sizes and physics assumptions; each simulation starts from initial conditions at the kinematic decoupling era ($z\sim 1059$), undergoes a discrete change from neutral gas to ionized gas thermal evolution at reionization ($z\sim 8$), and is finally processed into a Lyman-$\alpha$ forest transmitted flux cube. Streaming velocities suppress small-scale structure, leading to less violent relaxation after reionization. The changes in the gas distribution and temperature-density relation at low redshift are more subtle, due to the convergent temperature evolution in the ionized phase. The change in the BAO scale is estimated to be of the order of 0.12\%\ at $z=2.5$; some of the major uncertainties and avenues for future improvement are discussed. The predicted streaming velocity shift would be a subdominant but not negligible effect (of order $ 0.26\sigma$) for the upcoming DESI Lyman-$\alpha$ forest survey, and exceeds the cosmic variance floor.
\end{abstract}


\section{Introduction}
\label{sec:intro}

Modern cosmological observations have established a broad framework for the evolution of diffuse gas in the Universe. At early times, the cosmic gas was ionized and was tightly coupled, both thermally and kinetically, with the cosmic microwave background (CMB). At $z\sim 1100$, this gas recombined, rendering the Universe transparent; this epoch of cosmic recombination is directly measured by CMB anisotropy experiments, and the recombination dynamics and density and velocity perturbations at this epoch have been probed to high accuracy. The subsequent epoch -- the Dark Ages -- is not directly probed by observations, but structure formation theory predicts that the perturbations in the gas grew along with those of the dark matter. The dark matter perturbations went nonlinear, leading to the formation of dark matter halos, and sufficiently massive halos should have collected diffuse gas. Some of these halos hosted luminous objects that emitted ionizing photons, which are presumably responsible for the reionization of the Universe (currently estimated to have occurred at $z\sim 8$ based on CMB polarization data; \citealt{2016A&A...596A.108P}). Subsequently the diffuse gas in the intergalactic medium (IGM) was kept ionized and heated by the ionizing background radiation, with a small amount of H{\,\sc i} present in steady state due to ongoing recombinations. At lower redshifts, $z\lesssim 6$, the IGM can be probed using the H{\,\sc i} Lyman-$\alpha$ absorption forest against quasars. Large statistical samples of the Lyman-$\alpha$ forest are now available over the $2\lesssim z\lesssim 5$ range, and are broadly consistent with expectations from structure formation simulations (see \citealt{2016ARA&A..54..313M} for a review). The Lyman-$\alpha$ forest has also emerged as a key player in precision cosmology. In particular, the baryon-acoustic oscillation (BAO) scale has been measured in the Lyman-$\alpha$ forest (and in Lyman-$\alpha$-quasar cross-correlations) by the Baryon Oscillation Spectroscopic Survey (BOSS). This provides a ``standard ruler'' measurement of the rate of cosmic expansion at $z\sim 2.3$ \citep{2013A&A...552A..96B, 2013JCAP...04..026S, 2014JCAP...05..027F, 2015A&A...574A..59D, 2017arXiv170200176B}. This is currently the leading constraint in this redshift range, and its precision will improve in the future with experiments such as the Dark Energy Spectroscopic Instrument (DESI; \citealt{2016arXiv161100036D}).

Since the BAO scale is a localized feature in the correlation function rather than a broadband signal, it is generally considered to be a particularly robust cosmological probe. There are, however, a few types of astrophysical systematic errors that could couple directly to the BAO scale. One of these is the primordial {\em streaming velocity} $v_{\rm bc}$ of the baryons relative to the dark matter, whose correlations exhibit features at the BAO scale due to their common origin in acoustic waves in the primordial plasma. The streaming velocity modulates the formation of early baryonic structure since in regions of high $v_{\rm bc}$ the baryons do not fall into the smallest dark matter halos \citep{2010PhRvD..82h3520T}. If some imprint of these survives in low-redshift tracers of the density field -- whether galaxies or the Lyman-$\alpha$ forest -- then the BAO scale may be shifted, thereby representing a source of systematic error for cosmological measurements \citep{2010JCAP...11..007D, 2011JCAP...07..018Y, 2015MNRAS.448....9S, 2016PhRvL.116l1303B}. One's intuition is that this might be a very small effect: the structures most affected by streaming velocities are at very small scales, $\lesssim 10^7\,M_\odot$; such structures are below the Jeans mass after reionization and hence are destroyed. One might also expect that the most massive tracers used for BAO -- the luminous red galaxies (LRGs) -- would be the least affected \citep[e.g.][]{2016PhRvD..94f3508S}, and indeed there are already strong upper limits on streaming velocity effects on LRGs \citep{2013PhRvD..88j3520Y, 2016arXiv160706098S, 2017MNRAS.470.2723B}. Lyman-$\alpha$ clouds are much less massive, and one might expect them to show a stronger effect, though still small since the cloud masses are $\gg 10^7\,M_\odot$. The purpose of this paper is to present a first step toward quantifying these theoretical expectations.

The centerpiece of this paper is a series of hydrodynamic simulations of small boxes, intended to resolve structures down to the pre-reionization Jeans mass and then follow their imprint on the low-redshift IGM. Due to the simplified setups and limited simulation volumes, this paper should be viewed only as the first of many steps toward theoretical predictions for the Lyman-$\alpha$ forest BAO peak shift. We briefly describe here the physical ingredients that come into play, and the relation between past treatments of these ingredients and the present problem.

At the epoch of reionization, photoionization heats gas to a temperature of order $10^4$ K, with a corresponding increase in the Jeans mass. Since structures form hierarchically in the standard cold dark matter (CDM) paradigm, this means that reionization can actually destroy pre-existing small-scale baryonic structure. Any gas in shallow potential wells that is unable to cool on a sound-crossing time and not dense enough to self-shield will flow back out into the IGM. This ``evaporation'' process is a key event in cosmic history: a significant fraction of the baryonic matter in the Universe participated -- tens of per cents, if one includes the evaporation of filaments. It is also one of the few ways that the smallest structures in CDM cosmology affect the baryonic matter. However, probing the mini-halos is observationally very challenging: at the relevant redshifts ($z\gtrsim 7$), the Lyman-$\alpha$ forest is completely saturated, and even when the 21 cm fluctuations from the (pre-)reionization epoch are detected the mini-halo contribution \citep{2002ApJ...572L.123I} may be very difficult to disentangle from the diffuse contribution (see e.g.\ \citealt{2006ApJ...652..849F}). Therefore, much of the interest in mini-halos and their photoionization has been driven by their indirect effects on the reionization process itself. In particular, it is conceivable that mini-halos could act as photon sinks during reionization, with each atom undergoing multiple recombinations and photoionizations before escaping to the surrounding low-density IGM. This process has been studied analytically \citep{2001ApJ...551..599H, 2002ApJ...578....1B}, with numerical radiation-hydrodynamic simulations (\citealt{2004MNRAS.348..753S}; \citealt{2005MNRAS.361..405I}; \cchange{\citealt{2016ApJ...831...86P}}), and treated as sub-grid physics in models of reionization \citep{2006MNRAS.366..689C, 2014MNRAS.440.1662S}.

The destruction of small-scale structure also has a thermal and dynamical impact on the IGM. Hydrodynamic simulations have shown that by heating the IGM and raising the Jeans mass, reionization has a form of ``positive feedback'' that reduces the clumpiness and suppresses the rate of recombinations \citep[e.g.][]{2009MNRAS.394.1812P}. In order to understand the dynamical feedback, we recall that the re-heating of the IGM is sudden (in the sense that ionization fronts are highly supersonic in most regions), and that the re-heating temperature varies weakly with density. Thus immediately after reionization, the high-density regions (filaments and mini-halos) find themselves far out of pressure equilibrium with their low-density (and hence low-pressure, by the ideal gas law) surroundings. We therefore expect a violent relaxation of the IGM as it attempts to establish pressure equilibrium over a Jeans-scale patch. Indeed this happens in simulations of mini-halo evaporation: the mini-halo wind is preceded by a blast wave that propagates into the surrounding IGM \citep{2004MNRAS.348..753S}. Whether such perturbations leave any residual observable effects in the IGM at lower redshift is an open question. The heating and cooling processes in the diffuse IGM have an ``attractor'' solution, leading to a tight temperature-density relation at low redshifts with surprisingly little dependence on the initial conditions \citep{1997MNRAS.292...27H, 1998MNRAS.301..478T, 2015MNRAS.450.4081P, 2016MNRAS.456...47M}. Investigations of possible relics of hydrogen reionization bubbles in the temperature of the IGM -- and hence in the Lyman-$\alpha$ forest -- have therefore focused on the highest redshifts where saturation and statistics allow meaningful measurements, and are most sensitive for late reionization scenarios \citep{1994MNRAS.266..343M, 2003ApJ...596....9H, 2008ApJ...689L..81T, 2009ApJ...706L.164C, 2009ApJ...701...94F, 2014ApJ...788..175L, 2015MNRAS.447.3402B, 2015ApJ...813L..38D, 2016MNRAS.463.2335N, 2017ApJ...847...63O}.

However, none of these studies can be directly adapted to the problem of how the small-scale structures that are modulated by streaming velocities impact the low-$z$ Lyman-$\alpha$ forest. The detailed simulations by \citet{2005MNRAS.361..405I} consider a single spherically symmetric mini-halo with spherical infall, and so -- while they remain \cchange{a key} reference for mini-halos as photon sinks --  we cannot use them to study dynamical processes such as the channeling of blast wave energy into underdense regions, the effects of colliding blast waves, or the evolution of mini-voids. \cchange{\citet{2016ApJ...831...86P} track the evaporation of mini-haloes with geometry from cosmological initial conditions, but do not vary the streaming velocity or run their simulations to low redshift.} The cosmological simulations aimed at understanding the imprint of reionization bubbles on the Lyman-$\alpha$ forest require enormous volumes, and cannot resolve all structure down to the pre-reionization Jeans mass. The same is true of simulations that aim to reproduce the Lyman-$\alpha$ forest statistics used in precision cosmology \citep[e.g.][]{2014JCAP...07..005B,2015JCAP...12..017A} or IGM astrophysics \citep[e.g.][]{2017MNRAS.464..897B}. It is thus appropriate to revisit the outcome of small-scale structure disruption, focusing on intermediate scales where one can identify all the relevant baryonic structures, but where one can follow the propagation of shocks and the long-term thermal evolution of the disturbed gas.

This paper takes a minimalist approach to the physics, in the sense that it assumes that the smallest-scale structures leave no imprints on the low-redshift IGM other than via their photoionization-driven destruction. That is, we assume primordial star formation in low-mass haloes, and any resulting feedback, is negligible. Such feedback could in principle result in an additional shift in the BAO peak, of either sign. It would be much more difficult to compute and is thus far beyond the scope of the present work.

This paper is organized as follows. We begin by reviewing our conventions (\S\ref{sec:conventions}) and the order-of-magnitude physics of small-scale structure (\S\ref{sec:oom}). The methods for our simulations are discussed in \S\ref{sec:sim}. \S\ref{sec:results} describes the phenomenology observed in the simulations and the quantitative results for transparency of the IGM to the Lyman-$\alpha$ photons. In \S\ref{sec:implications}, we map the results into a change in the Lyman-$\alpha$ forest BAO scale. We discuss avenues for future work in \S\ref{sec:discussion}.

\section{Conventions}
\label{sec:conventions}

This paper assumes the background $\Lambda$CDM cosmology from the {\slshape Planck} 2015 ``$TT$+$TE$+$EE$+lowP+lensing+ext'' parameter set \citep{2016A&A...594A..13P}: $\Omega_{\rm b}h^2 = 0.02230$, $\Omega_{\rm m}h^2 = 0.14170$, $H_0=67.74$ km s$^{-1}$ Mpc$^{-1}$, $\sigma_8 = 0.8159$, and $n_s = 0.9667$. We consider a range of possible values for the redshift and duration of reionization, since the analysis in this paper is very sensitive to these parameters.

We write $T_4$ to denote a temperature in units of $10^4$ K, and $a_{-1}=10/(1+z)$ to denote the scale factor in units of 0.1 (convenient for reionization), and $\Delta = 1+\delta_{\rm b}$ to denote the gas density in units of the mean baryon density in the Universe. All masses and lengths in this paper are quoted without $h$ scalings (e.g.\ kpc, not $h^{-1}\,$kpc); the simulation code uses different units internally and these have been converted using the value of $h=0.6774$ in our background cosmology. We use ``ckpc'' and ``cMpc'' to denote comoving length units.

\section{Order-of-magnitude review of small-scale structure disruption}
\label{sec:oom}

Before proceeding to simulations, we revisit the basic orders of magnitude involved in small-scale structure disruption. For the purposes of setting our intuition, we consider mini-halo evaporation (see also \citealt{2005MNRAS.361..405I, 2016PhR...645....1B}), but it is important to remember that other structures (filaments and voids) also play a role, and are considered in a consistent way in the simulations. These scalings should be used as a reference point for the simulations (e.g.\ box size versus Jeans mass); should be compared to simulation results (e.g.\ timescales for kinetic energy injection); and motivate some prescriptions in the simulations (e.g.\ trapping overdensities and re-heating temperatures).

The number density of hydrogen nuclei is
\begin{equation}
n_{\rm H} = 1.9\times 10^{-4} a_{-1}^{-3} \Delta\,{\rm cm}^{-3},
\end{equation}
and the Hubble expansion time in the matter-dominated era is
\begin{equation}
t_H \equiv \frac1H = 820 a_{-1}^{3/2}\,{\rm Myr}.
\end{equation}
A halo of some mass $M$ has a physical virial radius of
\begin{equation}
R_{\rm v,phys} = 320 M_6^{1/3} a_{-1} \,{\rm pc},
\label{eq:Rphys}
\end{equation}
where $M_6$ is the halo mass in units of $10^6\,M_\odot$. The circular velocity at the virial radius is
\begin{equation}
V_{\rm c} = 3.7 M_6^{1/3} a_{-1}^{-1/2}\,{\rm km}\,{\rm s}^{-1}
\end{equation}
and the virial temperature is
\begin{equation}
T_{\rm v} = 1000 M_6^{2/3} a_{-1}^{-1}\,{\rm K}.
\end{equation}

Not all halos contain gas: in the pre-reionization era, thermal and ram pressure of the gas suppress the gas abundance in haloes below some filtering scale (e.g.\ \citealt{2007MNRAS.377..667N}), which is $\sim 2\times 10^5\,M_\odot$ depending on the baryon streaming velocity \citep{2011MNRAS.418..906T}. On the other hand, haloes at $M>3\times 10^7\,M_\odot$ have virial temperatures exceeding $10^4$ K, and can undergo Lyman-$\alpha$ cooling (and hence may be able to form stars, even in the presence of an H$_2$-dissociating UV background). The sterile gas-bearing mini-halos in between these masses -- i.e.\ at $0.2\lesssim M_6\lesssim 30$ -- are of direct interest to this paper. The \citet{1999MNRAS.308..119S} mass function places 8 per cent of the mass in halos of this range at $z=9$. Most of the remaining gas is diffuse, but not necessarily near mean density: by $z=9$, tens of per cents of the gas has already formed into filaments, which in turn are feeding the growth of the mini-halos.

An ionization front has a physical thickness
\begin{equation}
d_{\rm ion} = \frac1{n_{\rm H}\sigma_{\rm HI}} = 1400 a_{-1}^3 \Delta^{-1}\,{\rm pc},
\end{equation}
where we take $\sigma_{\rm HI} = 1.2\times 10^{-18}$ cm$^2$ at 24.6 eV just below the He{\,\sc i} ionization edge and we note the density dependence. Since gas in a halo is at least a factor of $\sim 100$ denser than cosmic mean, this means that the ionization front passing through a mini-halo will have a thickness that is a small fraction of the virial radius. As this ionization front passes into a mini-halo, it will become ``trapped'' as the dense, now-ionized gas attempts to recombine; the newly formed neutral atoms absorb incident ionizing photons and thus reduce their flux. The ionization front is predicted to trap when the rate of recombinations per unit area equals the incident photon flux:
\begin{equation}
\int_{r_{\rm t}}^\infty \alpha_{\rm B}n_{\rm H}^2\,dr = F.
\label{eq:trap}
\end{equation}
The details of the trapping depend on the radial profile of the gas in the halo. If one considers a singular isothermal sphere, the density profile is given by $\Delta = 60(r/R_{\rm v})^{-2}$, where the normalization gives an enclosed mean overdensity at $R_{\rm v}$ of 180. Substituting this into Eq.~(\ref{eq:trap}) predicts that trapping should occur at the radius $r_{\rm t}$ where:
\begin{equation}
F = 1200\alpha_{\rm B}\bar n_{\rm H}^2 \frac{R_{\rm v}^4}{r_{\rm t}^3},
\end{equation}
or
\begin{equation}
\frac{r_{\rm t}}{R_{\rm v}}
= \left( \frac{1200\alpha_{\rm B}\bar n_{\rm H}^2R_{\rm v}}{F} \right)^{1/3}
= 0.39 a_{-1}^{-5/3} M_6^{1/9} F_5^{-1/3},
\end{equation}
where $F_5$ is the incident ionizing photon flux in units of $10^5$ photons cm$^{-2}$ s$^{-1}$, and $\alpha_{\rm B} = 1.43\times 10^{-13}$ cm$^3$ s$^{-1}$ is the recombination coefficient of hydrogen at $T_4=2$ \citep{1991A&A...251..680P}. The trapping occurs at an overdensity of
\begin{equation}
\Delta_{\rm t} = 60 \left( \frac {r_{\rm t}}{R_{\rm v}} \right)^{-2} = 390 a_{-1}^{10/3} M_6^{-2/9} F_5^{2/3}.
\label{eq:trap-overdensity}
\end{equation}
It is seen that for the fiducial parameters, most of the gas in the halo is immediately ionized, but the trapping of the ionization front can leave behind a small neutral core. This is well inside the virial radius, and so the above estimates are likely to depend on the detailed gas structure inside the halo. The ionizing photon flux is related to the propagation speed of the ionization front,
\begin{equation}
v_{\rm i} = \frac{F}{n_{\rm H}(1+f_{\rm He})} = 4.9\times 10^3a_{-1}^3\Delta^{-1}\,{\rm km}\,{\rm s}^{-1},
\label{eq:vi}
\end{equation}
or in terms of the comoving distance traveled by the front per unit redshift:
\begin{equation}
\left|\frac{{\rm d}r_{\rm i}}{{\rm d}z}\right| = \frac{F}{n_{\rm H}(1+f_{\rm He})} t_{\rm H} = 4.1 a_{-1}^{9/2}\Delta^{-1}F_5\,{\rm cMpc}.
\end{equation}
We see that at $z\sim 9$, a flux of $F_5=1$ corresponds to an ionization front that propagates through 4 cMpc per $\Delta z=1$ at mean density, which is typical of reionization simulations. For the portions of the halo facing into the ionizing source at an oblique angle, the incident photon flux is of course less, and on the shaded side it may be much smaller.

The ionization of the main body of the halo heats the gas to a temperature of $T_4\sim 2$ \citep{1994MNRAS.266..343M, 2012MNRAS.426.1349M}, with a corresponding sound speed of $c_s = 21$ km s$^{-1}$. As long as the new temperature is well above the virial temperature, the gas component of the halo should explode on the sound-crossing time \citep{2001ApJ...551..599H}:
\begin{equation}
t_s = \frac{R_{\rm v}}{c_s} = 15 M_6^{1/3} a_{-1}\,{\rm Myr}.
\end{equation}
Simulations show that the cold, self-shielded neutral core is disrupted on a similar timescale as a shock wave detaches from the ionization front and plows through it \citep{2005MNRAS.361..405I}.
After the explosion, the halo gas emerges and drives a shock into the surrounding medium at a speed comparable to $c_s$. The gas will expand back to mean density when the expanding debris reaches $\sqrt[3]{180}$ times the virial radius. This occurs at a time of order
\begin{equation}
t_{\rm exp} = \sqrt[3]{180}\,t_s = 84 M_6^{1/3} a_{-1}\,{\rm Myr}.
\label{eq:texp}
\end{equation}
We expect that after time $t_{\rm exp}$, the emerging gas has collided with an amount of gas with total thermal energy comparable to the kinetic energy of the explosion (both energies are of $\sim Mc_s^2$). At times later than $\sim t_{\rm exp}$ after reionization, the kinetic energy of the initial explosion should thermalize in the IGM. Subsequently the ``conventional'' IGM thermal evolution should take over. The simulations in this paper show that $t_{\rm exp}$ gives a correct timescale for the thermalization of most of the kinetic energy, but that the IGM does not completely relax and that some motions and weak shocks survive $>1$ Gyr later.

The Jeans mass in the ionized IGM is
\begin{equation}
M_{\rm J} = \frac43\pi\left( \frac{\pi c_s}{\sqrt{4\pi G\bar\rho_{\rm m}}} \right)^3\bar\rho_{\rm m}
= 5.6\times 10^9 \left( a_{-1}T_4 \right)^{3/2}\,M_\odot.
\end{equation}
Structures in the IGM on scales smaller than the Jeans mass are no longer gravitationally bound, although given that the dynamical timescale is the age of the Universe they may persist for a cosmologically significant period of time.

\section{Simulation methodology}
\label{sec:sim}

We follow the formation of structure using a modified version of the smoothed particle hydrodynamics (SPH) code {\sc Gadget 2} \citep{2001NewA....6...79S, 2005MNRAS.364.1105S}. The simulation boxes followed are small, in order to properly resolve the formation and destruction of small-scale structure down to the Jeans mass of the cold, pre-reionization gas. Therefore, while cosmological initial conditions are used (in the sense of starting from a Gaussian random field and forming structures via gravitational instability), most boxes are not large enough to sample the linear regime at redshifts of interest to the Lyman-$\alpha$ forest. We therefore took a two-phase strategy to estimate the effects of the physics of interest on the Lyman-$\alpha$ forest. In ``Phase I,'' a common small box size (425 comoving kpc) and single set of initial conditions was used to explore the effects of numerical parameters and physical approximations. Based on this, a subset of parameter space was chosen for the ``Phase II'' simulations that explored a larger volume -- large enough to form multiple Lyman-$\alpha$ clouds -- and build up statistics.

All simulation boxes are initialized at the epoch of kinematic decoupling, $z_{\rm dec} = 1059$. They are then evolved using neutral gas physics until reionization. Reionization is treated as instantaneous, which should be valid as long as the ionization front is highly supersonic so that no hydrodynamic evolution occurs during its passage. At this instant of reionization $z_{\rm r}$, we reset the temperatures of the gas particles to account for the energy deposited by the ionization front. Thereafter, the simulation continues using heating and cooling terms appropriate to singly ionized primordial gas (H{\,\sc ii} + He{\,\sc ii}). The simulations do not currently incorporate He{\,\sc ii} reionization.

Some of the larger Phase II simulations were run on the Ruby cluster at the Ohio Supercomputer Center \citep{Ruby2015}.

\subsection{Parameters}

Each simulation has $2\times N^3$ particles, with equal numbers of gas and dark matter particles. No other particle types are used.

The gravitational softening length is set to $L/(5.6N)$, where $L$ is the comoving box size and hence $L/N$ is the initial comoving inter-particle spacing. This differs from the default choice of $L/(25N)$, which we find leads to spurious dynamical interactions between the gas and dark matter particles when run with initial conditions including a streaming velocity. If the gravitational softening length is too small, then there is a periodic interaction potential depending on the relative displacements of the dark matter and gas fields, and quantities such as the baryon kinetic energy undergo ``ripples'' as the gas flows over the grid of dark matter particles. This problem does not occur when the gas and dark matter particles are initialized with the same displacement and velocity field as in a standard simulation using {\sc Gadget 2} + {\sc N-GenIC}. It also does not appear to have been an issue for \cchange{some} other SPH simulations of the streaming velocity effects that were aimed at understanding early star formation (e.g.\ \citealt{2011ApJ...730L...1S, 2011MNRAS.412L..40M}), possibly due to a different choice of initial conditions. \cchange{A spurious inter-special coupling issue {\em was} noticed in \citet{2012ApJ...760....4O}; based on their Eq.~(B1), the squared ratio of the escape velocity from the dark matter particles to the RMS streaming velocity is
\begin{equation}
\frac{v_{\rm esc}^2}{\sigma_{\rm bc}^2} = 0.078 \left( \frac{M_{\rm p}}{10^3\,M_\odot}\right)^{2/3} \left( \frac{200}{1+z} \right) \left( \frac{0.18}{f_{\rm ms}} \right),
\end{equation}
where $M_{\rm p}$ is the particle mass and $f_{\rm ms}$ is the smoothing length in unit of the inter-particle spacing (0.04 default, 0.18 here). This is in accordance with our experience that if $f_{\rm ms}$ is too small, the streaming velocities are altered by the gravitational pull of individual particles. However, \citet{2012ApJ...760....4O} used glass-like initial conditions whereas we used a grid, so the phenomenology of the coupling is very different.}

The values of $L$ and $N$ are summarized in Table~\ref{tab:sim-param}. The Phase I simulations are based on a small box, roughly 1 post-reionization Jeans length on a side, that we use as a base to explore a wide variety of changes in the physics. The reference set of physics is used for subsequent larger boxes in Phase II.

\begin{table*}
\caption{\label{tab:sim-param}The parameters for our simulations. The ``reionization temperature'' column denotes either the uniform reionization temperature (cases with ``{\tt T}'') or a density-dependent reionization temperature according to a mean ionization front velocity and blackbody temperature of the incident photons (cases with ``{\tt P}''). The ``heating and cooling physics'' column describes deviations of the heating and cooling from the reference scenario. The letters denote: X=X-ray pre-heating; S=slow heating, post-reionization. The ``initial conditions'' are usually CDM (cold dark matter); the asterisk (*) denotes initial conditions with only the growing mode in the initial dark matter perturbations ($\nu_0=0$).}
\begin{tabular}{cccccccccc}
\hline\hline
Name & Comoving & Number of & DM particle & Gas particle & Streaming & Overdensity & Reionization & Heating \& & Initial \\
 & box size $L$ & particles & mass & mass & velocity & threshold  & parameters & cooling & conditions \\
 & [kpc] & & [$M_\odot$] & [$M_\odot$] & [km s$^{-1}$] & $\Delta_{\rm th}$ & $T_{\rm re}$ or $\bar v_{\rm i}, T_{\rm bb}$ & physics & \\
 & & & & & & & \!\!\! [Mm s$^{-1}$, $10^4$ K] \!\!\! & & \\
\hline
\multicolumn{10}c{Phase I Simulations} \\
\hline
Reference & \wsp425 & $2\times 128^3$ & $1.21\times 10^3$  & $2.27\times 10^2$ & 33 & 300 & {\tt T} 2.00 & ~ & CDM \\
nov & \wsp425 & $2\times 128^3$ & $1.21\times 10^3$  & $2.27\times 10^2$ & \wsp0 & 300 & {\tt T} 2.00 & ~ & CDM \\
pre & \wsp425 & $2\times 128^3$ & $1.21\times 10^3$  & $2.27\times 10^2$ & 33 & 300 & {\tt T} 2.00 & X & CDM \\
pre-nov & \wsp425 & $2\times 128^3$ & $1.21\times 10^3$  & $2.27\times 10^2$ & \wsp0 & 300 & {\tt T} 2.00 & X & CDM \\
D100 & \wsp425 & $2\times 128^3$ & $1.21\times 10^3$  & $2.27\times 10^2$ & 33 & 100 & {\tt T} 2.00 & ~ & CDM \\
D100-nov & \wsp425 & $2\times 128^3$ & $1.21\times 10^3$  & $2.27\times 10^2$ & \wsp0 & 100 & {\tt T} 2.00 & ~ & CDM \\
Soft & \wsp425 & $2\times 128^3$ & $1.21\times 10^3$  & $2.27\times 10^2$ & 33 & 300 & {\tt T} 2.00 & S & CDM \\
Soft-nov & \wsp425 & $2\times 128^3$ & $1.21\times 10^3$  & $2.27\times 10^2$ & \wsp0 & 300 & {\tt T} 2.00 & S & CDM \\
I5 & \wsp425 & $2\times 128^3$ & $1.21\times 10^3$  & $2.27\times 10^2$ & 33 & 300 & {\tt F} 5.00,5.00 & & CDM \\
I5-nov & \wsp425 & $2\times 128^3$ & $1.21\times 10^3$  & $2.27\times 10^2$ & \wsp0 & 300 & {\tt F} 5.00,5.00 & & CDM \\
Reference* & \wsp425 & $2\times 128^3$ & $1.21\times 10^3$  & $2.27\times 10^2$ & 33 & 300 & {\tt T} 2.00 & ~ & CDM* \\
nov* & \wsp425 & $2\times 128^3$ & $1.21\times 10^3$  & $2.27\times 10^2$ & \wsp0 & 300 & {\tt T} 2.00 & ~ & CDM* \\
\hline
LoRes1 & \wsp425 & $2\times \wsp96^3$ & $2.88\times 10^3$ & $5.38\times 10^2$ & 33 & 300 & {\tt T} 2.00 & ~ & CDM \\
LoRes1-nov & \wsp425 & $2\times \wsp96^3$ & $2.88\times 10^3$ & $5.38\times 10^2$ & \wsp0 & 300 & {\tt T} 2.00 & ~ & CDM \\
LoRes2 & \wsp425 & $2\times \wsp64^3$ & $9.72\times 10^3$ & $1.81\times 10^3$ & 33 & 300 & {\tt T} 2.00 & ~ & CDM \\
LoRes2-nov & \wsp425 & $2\times \wsp64^3$ & $9.72\times 10^3$ & $1.81\times 10^3$ & \wsp0 & 300 & {\tt T} 2.00 & ~ & CDM \\
\hline
\multicolumn{10}c{Phase II Simulations} \\
\hline
II-A & \wsp425 & $2\times \wsp64^3$ & $9.72\times 10^3$ & $1.81\times 10^3$ & 33 & 300 & {\tt T} 2.00 & ~ & CDM \\
II-An & \wsp425 & $2\times \wsp64^3$ & $9.72\times 10^3$ & $1.81\times 10^3$ & \wsp0 & 300 & {\tt T} 2.00 & ~ & CDM \\
II-B & \wsp850 & $2\times128^3$ & $9.72\times 10^3$ & $1.81\times 10^3$ & 33 & 300 & {\tt T} 2.00 & ~ & CDM \\
II-Bn & \wsp850 & $2\times128^3$ & $9.72\times 10^3$ & $1.81\times 10^3$ & \wsp0 & 300 & {\tt T} 2.00 & ~ & CDM \\
II-C & 1275 & $2\times192^3$ & $9.72\times 10^3$ & $1.81\times 10^3$ & 33 & 300 & {\tt T} 2.00 & ~ & CDM \\
II-Cn & 1275 & $2\times192^3$ & $9.72\times 10^3$ & $1.81\times 10^3$ & \wsp0 & 300 & {\tt T} 2.00 & ~ & CDM \\
II-D & 1701 & $2\times 256^3$ & $9.72\times 10^3$ & $1.81\times 10^3$ & 33 & 300 & {\tt T} 2.00 & ~ & CDM \\
II-Dn & 1701 & $2\times 256^3$ & $9.72\times 10^3$ & $1.81\times 10^3$ & \wsp0 & 300 & {\tt T} 2.00 & ~ & CDM \\
II-F & 2551 & $2\times 384^3$ & $9.72\times 10^3$ & $1.81\times 10^3$ & 33 & 300 & {\tt T} 2.00 & ~ & CDM \\
II-Fn & 2551 & $2\times 384^3$ & $9.72\times 10^3$ & $1.81\times 10^3$ & \wsp0 & 300 & {\tt T} 2.00 & ~ & CDM \\
\hline\hline
\end{tabular}
\end{table*}

\subsection{Initial conditions}

The gas is initialized to the CMB temperature, which is 2891 K at the epoch of kinematic decoupling. Particle positions and velocities are initialized using a modification of the built-in cosmological initial condition generator in {\sc Gadget 2}, {\sc N-Gen-IC}, \cchange{which uses the Zel'dovich approximation with the matter-dominated growing mode}. The principal modification is to place the initial perturbations in the dark matter only, i.e.\ the displacements and velocities of the gas particles are set to zero, while those of the dark matter are multiplied by $\Omega_{\rm m}/\Omega_{\rm c}$ (where $\Omega_{\rm c}=\Omega_{\rm m}-\Omega_{\rm b}$ is the density parameter for the dark matter). This is appropriate since until kinematic decoupling, the baryons have been locked to the radiation and hence are smoothly distributed on small scales.

We further excite the decaying as well as the growing modes in the initial conditions, by multiplying the dark matter particle displacements by a factor of $1+\nu_0$ and the velocities by a factor of $1-\frac32\nu_0$, where $\nu_0$ is the ratio of decaying to growing mode amplitude at the initial time. Based on the matter transfer functions from {\sc Class} v2.4.3 \citep{2011arXiv1104.2932L} at the initial scale factor $a_{\rm init} = 1/1060$ and at $5a_{\rm init}$, we compute this ratio as $\nu_0=0.21$ (at $k=1\,$Mpc$^{-1}$; this varies slowly with $k$ and is 0.22 at $k=100\,$Mpc$^{-1}$). This approach is only approximate because we are not treating the full dynamical effects of the radiation correctly (notably the radiation contribution to the Hubble expansion and the non-instantaneous decoupling of radiation from baryons). An alternative set of simulations (denoted with the asterisk, *) does not include the initial decaying mode, i.e.\ has $\nu_0=0$. This turns out to make only a small difference in the results.

We also allow for a relative streaming velocity ${\bmath v}_{\rm bc}$ between the gas and dark matter. The streaming velocity is coherent on scales of several cMpc, so we account for it by adding a velocity $(\Omega_{\rm c}/\Omega_{\rm m}){\bmath v}_{\rm bc}$ to the gas and $-(\Omega_{\rm b}/\Omega_{\rm m}){\bmath v}_{\rm bc}$ to the dark matter on the $x$-axis. The RMS value of the streaming velocity is 33 km s$^{-1}$ at $z_{\rm dec}$; our default simulations impose this RMS value.

\subsection{Heating and cooling}

The public version of {\sc Gadget 2} does not include heating or cooling terms, although these are straightforward to add.\footnote{They are added to the entropy evolution in {\tt hydra.c}.} Different heating and cooling physics applies to the neutral phase (pre-reionization) and the ionized phase (post-reionization). We include only the most important processes for gas of primordial composition.

In {\sc Gadget 2}, the thermal state of the gas is described by an entropy (and density); given the cosmological parameters, these can be converted into the more familiar temperature $T$ and $n_{\rm H}$ (the number density of hydrogen nuclei). The ratio of helium to hydrogen nuclei by number, $f_{\rm He} = 0.079$, is a constant. All heating and cooling rates here are written in terms of
\begin{equation}
\frac{\dot A_i}{A} \equiv \frac{2\dot E_i}{3nk_{\rm B}T},
\end{equation}
where $A$ is the particle ``entropy'' in the code ($A=p\rho^{-5/3}$, where $p$ is pressure and $\rho$ is density), $k_{\rm B}$ is Boltzmann's constant, $n$ is the total number density of particles (including atoms in the neutral case, or ions + electrons in the ionized case), and $\dot E_i$ is the net volumetric heating rate in erg cm$^{-3}$ s$^{-1}$ from process $i$. The quantity $\dot A_i/A$ has units of s$^{-1}$ and should be thought of as the fractional rate of increase of thermal energy from the relevant process.

\subsubsection{Neutral gas}

For the neutral phase, the only heating and cooling source in our Reference simulation is Compton scattering from the CMB:
\begin{equation}
\frac{\dot A_{\rm Compton}}A = \Gamma_{\rm C,n} \left( -1 + \frac{T_\gamma}T \right),
\end{equation}
where $T_\gamma = 2.725(1+z)$ K is the CMB temperature, the second term accounts for Compton heating from the CMB, and the Compton cooling rate \citep[e.g.][]{1965PhFl....8.2112W} for neutral gas is
\begin{equation}
\Gamma_{\rm C,n} = \frac{8\sigma_{\rm T}a_{\rm R}T_\gamma^4}{3m_ec} \frac{n_e}{n} = \frac{3.494}{t_H} a_{-1}^{-5/2} \frac{2x_e}{1.079+x_e},
\end{equation}
where $\sigma_{\rm T}$ is the Thomson cross section, $a_{\rm R}$ is the radiation density constant, $m_e$ is the electron mass, $c$ is the speed of light, and $t_H$ is the Hubble time. The second equality makes use of the {\slshape Planck} cosmological parameters and assumes matter domination. Here $x_e$ was obtained by fitting {\sc Hyrec} \citep{2011PhRvD..83d3513A} output with a 9th-order polynomial fit of $\ln x_e$ vs. $z$.

The Reference simulation does not contain any provision for cooling. In fact, even a small box that will contain 1 Jeans mass of gas at $z\sim 3$ is likely to contain at least one $T_{\rm vir}>10^4$ K halo at $z\sim 8$, which in reality will undergo Lyman-$\alpha$ cooling. By artificially turning this process off, the Reference simulation ignores the possibility of star formation or feedback from this halo. Smaller halos can cool via H$_2$ lines, but this process may be suppressed by the negative feedback from the non-ionizing ($h\nu<13.6\,$eV) ultraviolet continuum from early stars, which can dissociate H$_2$ \citep{1997ApJ...476..458H}. The Reference simulation assumes that the H$_2$-dissociating feedback is strong, as suggested by theoretical models \citep[e.g.][]{2012MNRAS.419..718H}.

Similarly, the Reference simulation does not include pre-heating of the neutral gas by penetrating, high-energy radiation. This possibility is explored in a variation on the reference scenario (\S\ref{ss:X-ray}).

\subsubsection{Reionization}

The reionization process itself affects the temperature through photo-ionization heating, Lyman-$\alpha$ cooling, and changes in the number of degrees of freedom. Studies have suggested a post-reionization temperature $T_{\rm re}$ of order $2 \times 10^4$ K, depending on the ionizing photon spectrum and the velocity of the ionization front \citep{1994MNRAS.266..343M, 2012MNRAS.426.1349M}. In this paper, we implement two versions of the reionization temperature:
\begin{list}{$\bullet$}{}
\item A uniform post-reionization temperature (denoted ``{\tt T}'' in Table~\ref{tab:sim-param}). This is the default, with value $T_{4,\rm re}=2.00$.
\item A density-dependent post-reionization temperature coming from a simple ionization front model (denoted ``{\tt F}'' in Table~\ref{tab:sim-param}), as described in Appendix~\ref{app:I-front}. In this model, the higher-density regions reionize to a lower temperature than the lower-density regions because of the greater importance of Lyman-$\alpha$ cooling in the partially ionized phase. The model has two parameters: (i) the ionization front velocity at mean density $\bar v_{\rm i}$ (note $v_{\rm i}\propto \Delta^{-1}$) and (ii) the spectrum of ionizing photons driving the front, here taken to be a blackbody of temperature $T_{\rm bb}$ truncated at 4 Ry (the He{\sc\,ii} ionization edge).\footnote{To save time, the model of Appendix~\ref{app:I-front} was run on a grid of velocities from $6.5 \le \log_{10} v_{\rm i} \le 9.5$ ($v_{\rm i}$ in cm s$^{-1}$), and a cubic polynomial was fit to the resulting temperature. Velocities outside the range of validity of the fit are replaced with $\log_{10} v_{\rm i} = 6.5$ or 9.5, respectively.}
\end{list}

Since {\sc Gadget 2} is a hydrodynamics code, and we have not implemented a radiative transfer module on top of it, the self-shielding of dense gas must be added via some prescription. Our ``default'' procedure is to turn off heating and cooling at densities above some overdensity threshold $\Delta_{\rm th}$ (default: $\Delta_{\rm th} = 300$). Note that the turning off of heating and cooling in dense gas also prevents the well-known catastrophic cooling of gas in halos that form at late times and have sufficiently high virial temperature. If no preventative measures are taken, {\sc Gadget 2} takes extremely short time steps in order to track these regions, which have no effect on the Lyman-$\alpha$ forest.

\subsubsection{Ionized gas}

For the ionized phase (post-reionization), the Compton process is again relevant, but now $n_e/n=0.5$, and the Compton heating from the CMB can be neglected as $T_\gamma\ll T$:
\begin{equation}
\frac{\dot A_{\rm Compton}}A = -\frac{3.494}{t_H} a_{-1}^{-5/2}.
\end{equation}
Additionally, one includes the recombination cooling and photo-ionization heating from H{\,\sc i} and He{\,\sc i}. For H{\,\sc i},
\begin{equation}
\frac{\dot A_{\rm H}}A = \frac13 \alpha_{\rm A} n_{\rm H} \left( -\frac32 + c_1 + \frac{\langle\epsilon_{\rm HI}\rangle}{k_{\rm B}T} \right),
\end{equation}
where $c_1 = -d\ln\alpha_{\rm A}/d\ln T$ is the power-law index of the recombination coefficient, and $\langle\epsilon_{\rm HI}\rangle$ is the mean energy of the photo-electron produced by ionizing an H{\,\sc i} atom. The same relation applies to He, but with the replacement of the recombination coefficients and with $n_{\rm H}\rightarrow n_{\rm He}=0.079n_{\rm H}$. We use the \citet{1991A&A...251..680P} fits to the Case A recombination rates,
\begin{eqnarray}
\alpha_{\rm A}({\rm H}) &=& \frac{5.596\times 10^{-13}T_4^{-0.6038}}{1+0.3436T_4^{0.4479}}\,{\rm cm}^{3}\,{\rm s}^{-1}
~~{\rm and} \nonumber \\
\alpha_{\rm A}({\rm He}) &=& \frac{8.295\times 10^{-13}T_4^{-0.5606}}{1+0.9164T_4^{0.2667}}\,{\rm cm}^3\,{\rm s}^{-1}.
\end{eqnarray}
The energy input per photo-ionization is taken from the \citet{2012ApJ...746..125H} model at $z=6$: $\langle \epsilon_{\rm HI} \rangle = 4.2$ eV and $\langle \epsilon_{\rm HeI}\rangle=7.2$ eV. These can vary with redshift as the spectral shape of the UV background changes, but for simplicity we do not include such changes; \citet{2012ApJ...746..125H} find only a 6 per cent change from $z=6$ to $z=2.5$.

For free-free cooling, we include the fitting formula from \citet[Eq. 10.10]{2011piim.book.....D}:
\begin{equation}
\frac{\dot A_{\rm f-f}}A = -4.97\times 10^{-14} n_{\rm H} T_4^{-0.45}\,{\rm s}^{-1}.
\end{equation}
Finally, at high temperatures, He{\,\sc ii} line cooling becomes important. Such temperatures are in fact reached in the photo-ionization driven winds from minihaloes. We include the excitations of He{\,\sc ii} to $n=2$ and $n=3$, following the collision strengths of \citet{1987MNRAS.224..801H}; the resulting function can be fit with an error of $<2$ per cent by the function
\begin{equation}
\frac{\dot A_{\rm line}}A = -2.336\times 10^{-8}\frac{n_{\rm H}}{T_4^{3/2}}(1+0.0663T_4){\rm e}^{-47.34/T_4} \,{\rm s}^{-1}
\label{eq:Aline}
\end{equation}
over the range $T_4<10$. Note the exponential cutoff due to the finite energy ($k_{\rm B}\times 4.734\times 10^5$ K) of the first excited state ($n=2$) of He{\,\sc ii}. This function should in principle be multiplied by the fraction of He in the He{\,\sc ii} ionization state, here assumed to be 1 since we are not including He{\,\sc ii} reionization.\footnote{This assumption fails when gas falls into massive halos, shock-heats to temperatures $>7\times 10^4$ K, and the He{\,\sc ii} is collisionally ionized to He{\,\sc iii}; the line cooling is then less than predicted by Eq.~(\ref{eq:Aline}). The affected regions are extremely overdense and would be saturated in the Lyman-$\alpha$ forest.}

In order to prevent stiff equation behaviour at high density, all of the density-squared processes (i.e.\ all except Compton cooling) are suppressed by a factor of $2000/\alpha_{\rm A}({\rm H})n_{\rm H} t_H$ when $\alpha_{\rm A}({\rm H})n_{\rm H} t_H>2000$ -- that is, when the recombination time becomes less than 1/2000 of the Hubble time. The density-squared processes lead to net heating for cool gas, and net cooling for hot gas; the transition temperature is $T_{\rm cr}=3.9\times 10^4$ K, and in the absence of other processes sufficiently dense gas with this heating and cooling model would be driven to a temperature $T=T_{\rm cr}$. A homogeneous IGM with the thermal evolution terms described above would always have $T<T_{\rm cr}$, but shocks can drive $T>T_{\rm cr}$ and make He{\,\sc ii} line cooling relevant. Because of the disassembly of structures, these shocks can even affect gas that ends up near mean density at $z\sim 3$.

A test case was run for a nearly homogeneous universe ($\sigma_8$ set to 0.001) with $2\times 64^3$ particles and reionization at $z_{\rm re}=8$ to verify that the thermal evolution in the modified {\sc Gadget-2} agrees with the direct solution of the ODEs. The maximum error is 1.2 per cent.

{\sc Gadget 2} does not include thermal conduction, and we do not add it here. Using the \citet{1953PhRv...89..977S} conductivity, the comoving thermal diffusion length is
\begin{equation}
\frac1a \sqrt{\frac{K t_{\rm H}}{c_p}} = 0.11a_{-1}^{5/4}T_4^{5/4}\Delta^{-1/2}\,{\rm ckpc},
\end{equation}
where $K$ is the thermal conductivity (units: erg$\,$cm$^{-1}\,$K$^{-1}$), $c_p$ is the heat capacity at constant pressure per unit volume (units: erg$\,$K$^{-1}\,$cm$^{-3}$), and $t_{\rm H}$ is the Hubble time. This is small compared to the particle size in our simulations.

After reionization, Compton drag once again acts on the baryons and is expected to induce a relative velocity between baryons and CDM of the same order of magnitude as the primordial component; see \citet{2017arXiv170507843S} for a detailed investigation. We have not included this effect in our simulations, since it has different spatial dependence from the primordial streaming velocities (i.e.\ it corresponds to a different biasing term than $b_v$) and is not expected to shift the BAO peak \citep{2017arXiv170507843S}.

\subsection{Lyman-$\alpha$ transmission}
\label{ss:Lya-trans}

In the small Phase I boxes, it is not possible to extract a true Lyman-$\alpha$ ``forest:'' the 425 ckpc size of the Reference box is comparable to the Jeans scale, and to the size of the smallest features in the forest. An alternative way to see this is that it corresponds to only 31 km s$^{-1}$ at $z=2.5$, which is only $1.5$ times the thermal width of the Lyman-$\alpha$ line at $10^4$ K. Nevertheless, it is of interest as an order-of-magnitude guide to when the various treatments of small-scale structures in this paper are relevant to Lyman-$\alpha$ transmission. The Phase II simulations have larger boxes, up through 2.55 cMpc, or 184 km s$^{-1}$ at $z=2.5$.

Lyman-$\alpha$ transmission is computed by choosing one of the three axes of the box to be the line of sight direction (all final statistics are averaged over the three axes). Gas particles are then assigned an H{\,\sc i} abundance proportional to $\Delta^2\alpha_{\rm A}(T)$, as appropriate for a mostly ionized plasma in a uniform ionizing background. A cutoff is applied at $\Delta_{{\rm th,Ly}\alpha}=100$ (i.e.\ particles at overdensities $\Delta>\Delta_{{\rm th,Ly}\alpha}$ are excluded from the optical depth cube, and particles at overdensities $\Delta>0.8\Delta_{{\rm th,Ly}\alpha}$ are downweighted by a factor that linearly interpolates between 1 at $0.8\Delta_{{\rm th,Ly}\alpha}$ and 0 at $\Delta_{{\rm th,Ly}\alpha}$), although we have checked that the main results in this paper are not sensitive to the cutoff. The H{\,\sc i} is then interpolated onto an $N_{\rm cell}\times N_{\rm cell}\times N_{\rm cell}$ grid (default: $N_{\rm cell}=32$), based on the redshift-space positions of the particles. In the two transverse directions, interpolation is performed linearly. In the line-of-sight direction, the particles are smoothed with a Gaussian at the thermal width corresponding to the temperature $T$ and the mass of the hydrogen atom. Because of the small box size, the Gaussian is allowed to ``wrap'' in the sense that we include the periodic-box images in building the Lyman-$\alpha$ forest cube. The intrinsic smoothing length of the gas particles (i.e.\ the smoothing length defined by the SPH routines in {\sc Gadget} 2) is smaller than the grid size and is not explicitly taken into account.

The aforementioned procedure generates an opacity cube -- i.e.\ a map of $\tau$ -- but with arbitrary normalization. The normalization is set by the requirement to reproduce the observed mean flux $\langle F\rangle = \langle{\rm e}^{-\tau}\rangle$ of the Lyman-$\alpha$ forest, fit by $\langle F \rangle = \exp(-0.0023a^{-3.65})$ at $1.7<z<4$ \citep{2007MNRAS.382.1657K}. We then report the required normalization in the form of $\tau_1$, the optical depth of a mean-density patch of gas at the mean expansion rate and $T_4=1$.

Variations with respect to changes in $N_{\rm cell}$ ($32\rightarrow 64$) and $\Delta_{{\rm th,Ly}\alpha}$ ($100\rightarrow 300$) are shown in Table~\ref{tab:transparent}.

\subsection{Variations}

We now turn to the variations in heating and cooling physics.

\subsubsection{X-ray preheating}
\label{ss:X-ray}

Since heating by Lyman-$\alpha$ photons is inefficient due to radiative transfer effects \citep{2004ApJ...602....1C}, the most likely source of pre-reionization heating was X-ray radiation \citep{1997ApJ...475..429M}, which can penetrate deep into neutral material and then undergo a photoelectric absorption and thermalize the energy of the photo-electron. The amount of such X-ray heating is highly uncertain, with models for the IGM temperature at $z=10$ ranging from $\sim 30$ K to many hundreds of K and with a number of revisions as the theory has developed \citep[e.g.][]{2014Natur.506..197F}. The X-ray heating rate $\dot T(t)$ depends on the X-ray background but not on the density\footnote{We do not track X-ray ionization, nor positive feedback due to the increased thermalization efficiency of subsequent X-rays in a partially ionized medium.}; our simulation with X-rays (``X'' in Table~\ref{tab:sim-param}) has a steeply rising heating rate $\dot T(t)\propto a^5$, and is normalized to heat mean-density gas by $\Delta T = 300$ K by $z=9$. In this scenario the gas temperature rises above the CMB temperature at $z=16$, consistent with the more extreme scenarios in \citet{2014Natur.506..197F}; see the blue-dashed curves in their Figure 2. Physically, one expects that this ``pre-heating'' of the IGM might puff up some structures before an ionization front arrives, in which case the dynamical relaxation effects following reionization might be reduced.

\subsubsection{Heating of dense gas}
\label{ss:slow}

The default model for handling dense gas in these simulations is to set the temperature at $z=z_{\rm re}$ from the usual reionization prescription, but to turn off (non-adiabatic) heating and cooling terms. An alternative is the ``slow heating'' prescription, in which material at $\Delta>\Delta_{\rm th}$ starts cold, at 10 times lower temperature than given by the reionization temperature prescription. It is then subjected to an artificial heating prescription, given by $\dot T_4 = 0$ if $T_4>1$ and $\dot T_4 = (50\,{\rm Myr})^{-1}$ if $T_4<1$. This prescription heats gas to $10^4$ K on a timescale of 50 Myr, i.e.\ roughly the mini-halo evaporation time expected from analytic arguments or from the \citet{2005MNRAS.361..405I} simulations.

Neither the default nor the slow-heating model is a realistic representation of the microphysics of halo evaporation. Rather, the two models -- as well as changing the threshold $\Delta_{\rm th}$ -- should be thought of as alternative phenomenological prescriptions to control the rate of disassembly of the mini-halos. Fortunately, the total amount of IGM gas in these very dense regions is small, and the major results of this paper do not depend on the prescription chosen.

\section{Simulation results}
\label{sec:results}

We now turn to the results of the Phase I simulations. We begin with a phenomenological description of the simulation results for the ``reference'' scenario, before considering the effects of alternatives. The Phase II simulations are considered at the very end (\S\ref{ss:phase2}).

\subsection{The reference simulations}

\begin{figure}
\includegraphics[width=3.2in]{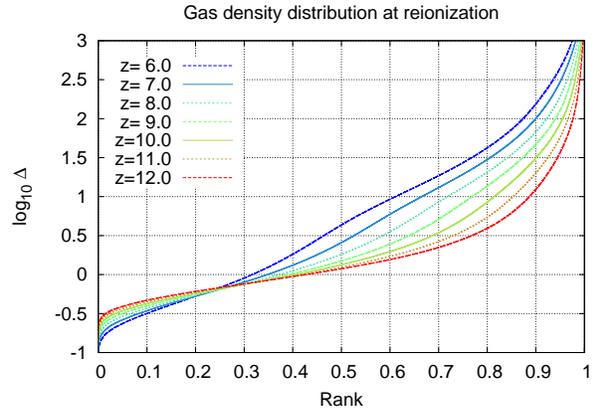}
\caption{\label{fig:dplot}The density distribution of the cold gas at reionization in a reference box ($L=425$ kpc), for a range of possible reionization redshifts. The gas particles at each redshift were rank-ordered by density; the fractional rank is shown on the horizontal axis, and $\log_{10}\Delta$ is shown on the vertical axis. At $z=8$ -- the currently favoured redshift of reionization -- we find that 28 per cent of the gas is at $\Delta>10$ and 8 per cent at $\Delta>100$, while 20 per cent is at $\Delta<0.5$.}
\end{figure}

\begin{figure*}
\includegraphics[width=6.2in]{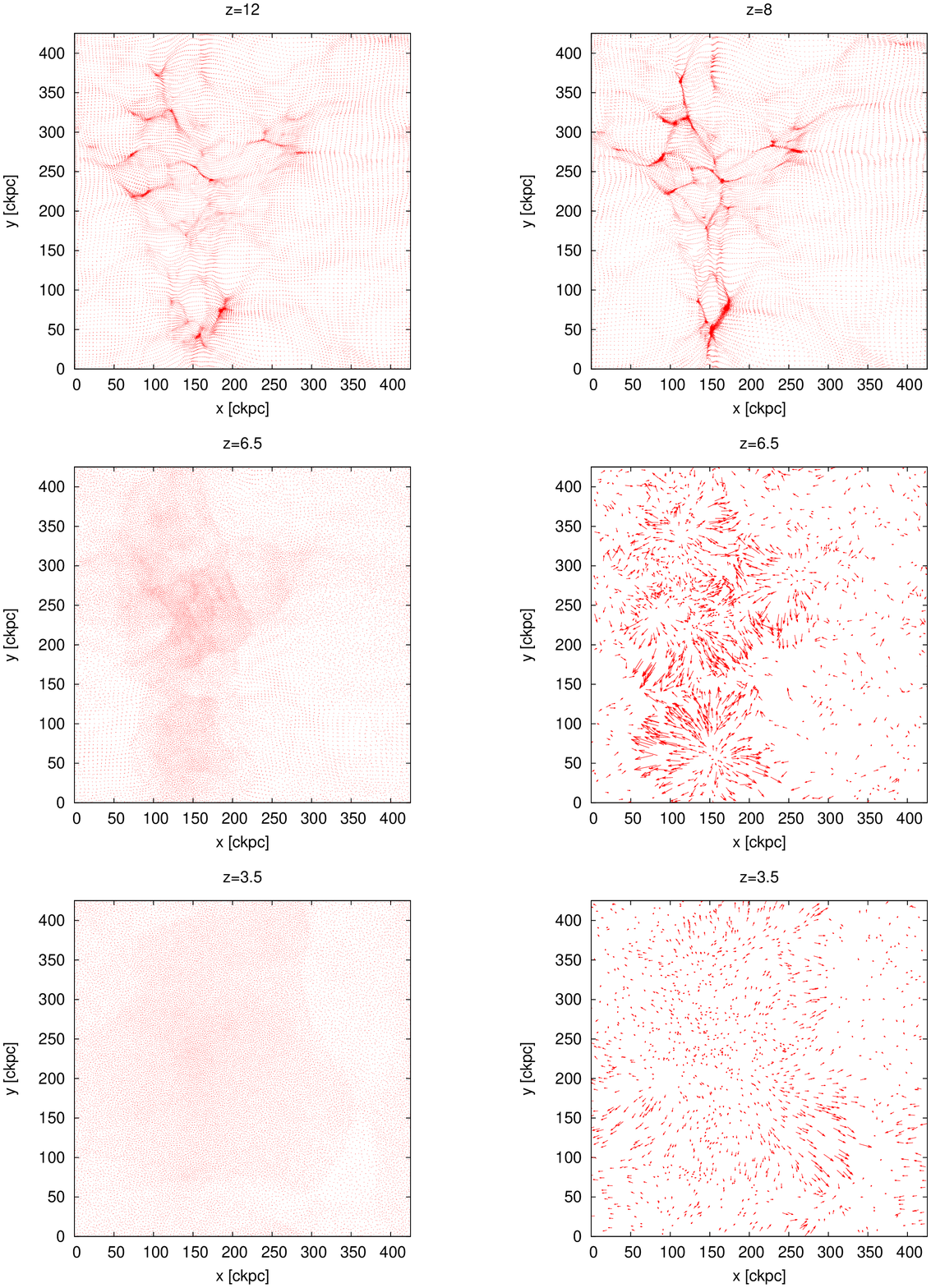}
\caption{\label{fig:hot-dens}A 5.9 ckpc thick slice of the gas particles in the Reference simulation, in the box that reionizes at $z_{\rm re}=8$. The simulation $z$-coordinate range (vertical dimension, not redshift) is 0---6 ckpc. {\em Top row}: The neutral gas distribution at $z=12$ (top left) and $z=8$ (top right). {\em Middle row}: The same slice at $z=6.5$. The gas density is shown in the left panel, and the velocity distribution in the right panel. Particles are randomly down-sampled by a factor of 20, with the vectors shown with the scale that 1 km s$^{-1}$ equates to 1 ckpc. {\em Bottom row}: Same at $z=3.5$.}
\end{figure*}

In Figure~\ref{fig:dplot}, we show the density distribution of cold gas at reionization in the ``reference'' 425 kpc box. The fraction of the gas mass at $>10\times$ mean density rises from 11 per cent at $z=12$ up to 39 per cent at $z=6$. The density distribution is often summarized in terms of a ``clumping factor''
\begin{equation}
C = \frac{\langle n^2\rangle}{\langle n \rangle^2} = \int_0^\infty \Delta^2P(\Delta)\,{\rm d}\Delta,
\end{equation}
which is the fractional variance of the density perturbations. It is $1$ for a homogeneous universe. We consider here $C_{100}$ where the subscript 100 indicates that the variance computation is cut off at 100 times the mean density, thereby excluding the interiors of halos. The Reference box has $C_{100}=6$ at $z=10$, rising to 10 at $z=6$. (For comparison, using a somewhat larger box and higher resolution, \citealt{2013ApJ...763..146E} find $C_{100}=15$ at $z=6$.) A slice through the gas distribution is shown in the top row of Figure~\ref{fig:hot-dens}. It should be evident that the gas is distributed very inhomogeneously on small scales.

Prior to reionization, most of the gas follows the adiabatic cooling relation, $T\sim 0.02(1+z)^2\Delta^{2/3}\,$K \citep{2009MNRAS.397..445S}, but -- especially at high densities -- the gas is significantly heated by shocks. We find that at $z=8$, the temperature at mean density has fallen to 2 K, and some gas has fallen to 0.5 K. However 10 per cent of the gas is at temperatures exceeding 540 K, 1 per cent is at $T>8900$ K, and 0.1 per cent at $T>2.1\times 10^4$ K. This is expected, given that in the reference box the median mass of the largest halo (predicted by the \citealt{1999MNRAS.308..119S} mass function) is $7\times 10^7\,M_\odot$, containing 2 per cent of the mass and with a virial temperature of $1.5\times 10^4$ K.

\begin{figure}
\includegraphics[width=3.2in]{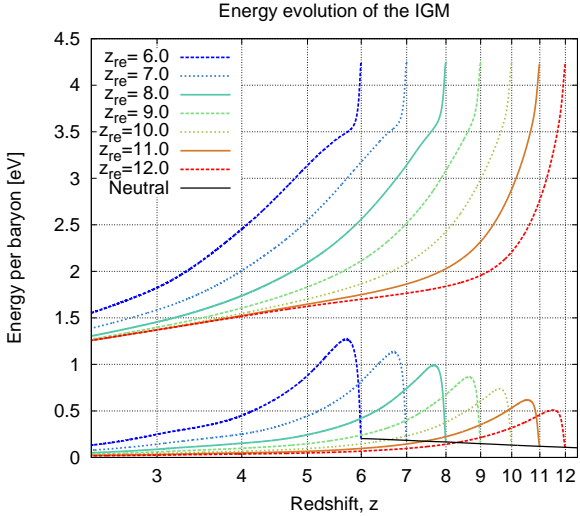}
\caption{\label{fig:energy}The energy per baryon in the Reference simulation. The black curve at bottom shows the kinetic energy in the simulation run using the neutral gas. The coloured lines show the kinetic (bottom) and thermal (top) energies after reionization.}
\end{figure}

After reionization, the small-scale structure in the Reference simulation rapidly disassembles. The energetics of the disassembly are shown in Figure~\ref{fig:energy}. The outflows from dense structures are seen in the gas kinetic energy, which typically rises on a timescale of $\sim 20$ Myr following reionization. On a corresponding timescale, the thermal energy rapidly drops, as the thermal energy in dense structures is tapped to power the outflows. The kinetic energy of outflows is a significant fraction of the total energy budget of the diffuse baryons; for $z_{\rm re}=7$, it is $\sim 20$ per cent. A significant decline in the kinetic energy is seen at $\gtrsim 100$ Myr, as the outflows collide with surrounding gas and convert their kinetic energy back into thermal energy. Thereafter, the kinetic energy declines toward zero,\footnote{Some of the decline is due to Hubble friction, but even the kinetic energy divided by $(1+z)^2$ is declining, indicating that kinetic energy is indeed being converted back to thermal energy.} and the thermal energy curves converge for different reionization redshifts, consistent with a sub-Jeans-mass patch of gas approaching pressure equilibrium and moving onto the attractor temperature-density relation.

We parameterize the extent of the small-scale structure disassembly by three parameters. First is the kinetic energy injection per baryon, $\epsilon_{\rm k}$, defined by the maximum kinetic energy per baryon (the peak of the curves in Figure~\ref{fig:energy}) minus the kinetic energy per baryon at the instant of reionization. The second is the rise time, $t_{\rm rise}$, from the instant of reionization to the peak kinetic energy per baryon. Finally, there is the fall time, $t_{\rm fall}$, which is the time for the kinetic energy per baryon to fall from its peak value to half of its peak value. We find, in the Reference simulation, at $z_{\rm re} = 12,\ 10,\ 8,\ 6$ that the kinetic energy injection is $\epsilon_{\rm k} = 0.40,\ 0.61,\ 0.83,\ 1.07\, $eV baryon$^{-1}$; the rise time is $t_{\rm rise} = 22,\ 26,\ 35,\ 63$ Myr; and the fall time is $t_{\rm fall} = 124,\ 146,\ 201,\ 343$ Myr. The rise and fall times can be compared to the order-of-magnitude estimates $t_s$ and $t_{\rm exp}$ of \S\ref{sec:oom}, which are seen to be good at the factor of 2--3 level.

\begin{figure*}
\includegraphics[angle=-90,width=6.9in]{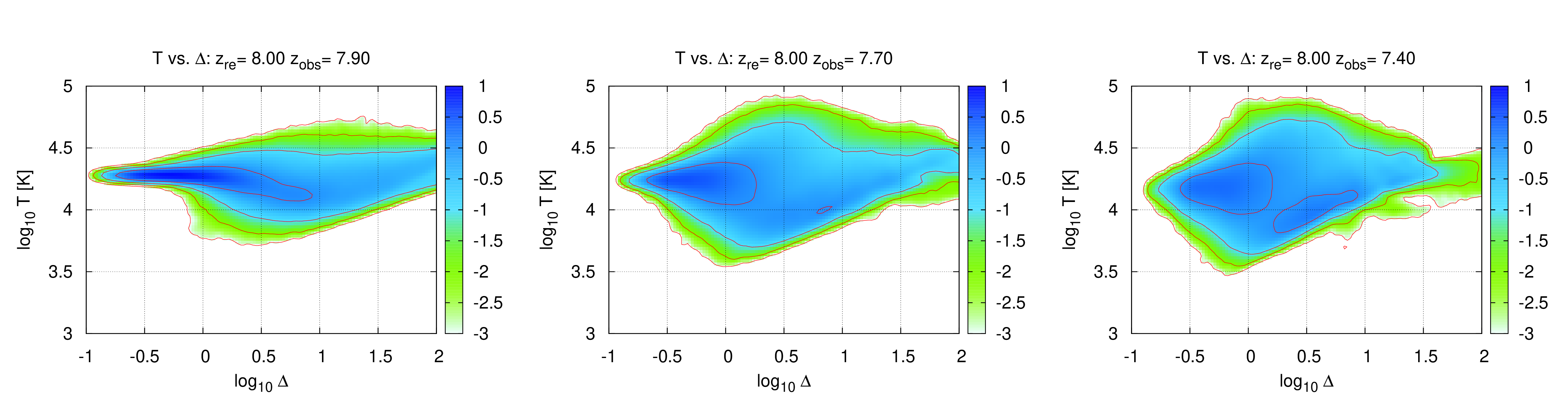}
\vskip-0.25in
\includegraphics[angle=-90,width=6.9in]{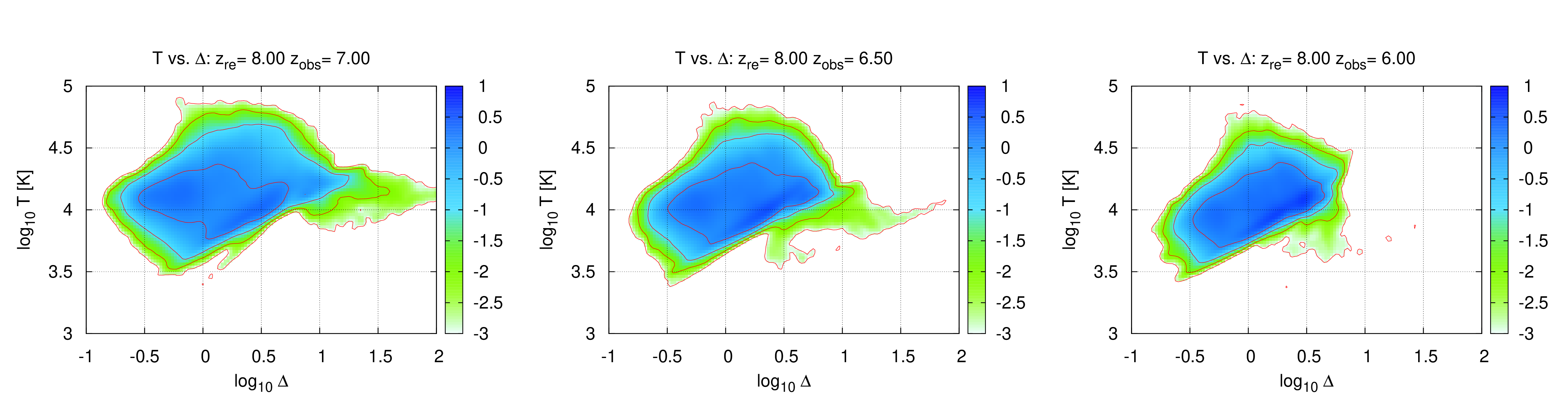}
\caption{\label{fig:early}The early evolution of the temperature-density relation in the $z_{\rm re}=8$ simulation.The particles have been smoothed with a Gaussian kernel density estimator (KDE) with a full-width at half maximum of 0.05 dex on each axis. The colour scale indicates the mass-weighted distribution of the gas, in units of log$_{10}$ probability per decade in temperature per decade in density. That is, a color of $-2$ (green) indicates that ${\rm d}P/{\rm d}\log_{10}T\,{\rm d}\log_{10}\Delta = 0.01$. Contours are shown for every decade in probability density.}
\end{figure*}

The temperature-density relation during the immediate post-reionization era is shown in Figure~\ref{fig:early}. In the first panel ($z=7.9$), 10 Myr after reionization, one can still see the initial temperature of the reionized gas ($\log_{10}T_{\rm re} = 4.29$). Material in the densest regions is starting to expand and adiabatically cool, hence the dip in the temperature-density relation at $\log_{10}\Delta\sim 1$. At still higher densities, the gas is hotter again: the number of recombinations per hydrogen atom in 10 Myr is $\Delta/130$, so the gas near the right edge of the plot is undergoing significant photo-heating. In the second panel ($z=7.7$), 33 Myr after reionization, we see the development of a {\em bimodal} temperature-density relation. First, there is a low-entropy sequence, consisting of former mini-halo and mini-filament gas that is undergoing a combination of adiabatic expansion and photo-ionization heating, leading to a conventional power-law temperature-density relation. There is also a high-entropy cloud of mini-void gas; some remains at low density, but by $z=7.7$, some of this gas has been compressed and heated to temperatures exceeding $3\times 10^4$ K. In some small regions, this compression is particularly violent: 1.5 per cent of the gas particles are heated to $4\times 10^4$ K, and the hottest particle in the simulation is at $9.3\times 10^4$ K. Inspection of the locations of these extremely hot particles shows them to be the result of a shock breakout from the mini-halos into the surrounding mean-density gas; however these particles rapidly cool and it is unlikely that the extreme temperatures are relevant for the later evolution of the system. In the third panel ($z=7.4$), 69 Myr after reionization, the bimodal temperature-density relation is fully formed, and it is clearly seen that the low-entropy sequence is being photo-heated to a higher adiabat. The subsequent evolution (bottom panel of Fig.~\ref{fig:early}) shows that the high-entropy cloud begins to descend toward the low-entropy sequence. It also shows that the density distribution of the gas is narrowing as gas pressure smears out the small-scale density perturbations.

\begin{figure*}
\includegraphics[angle=-90,width=6.9in]{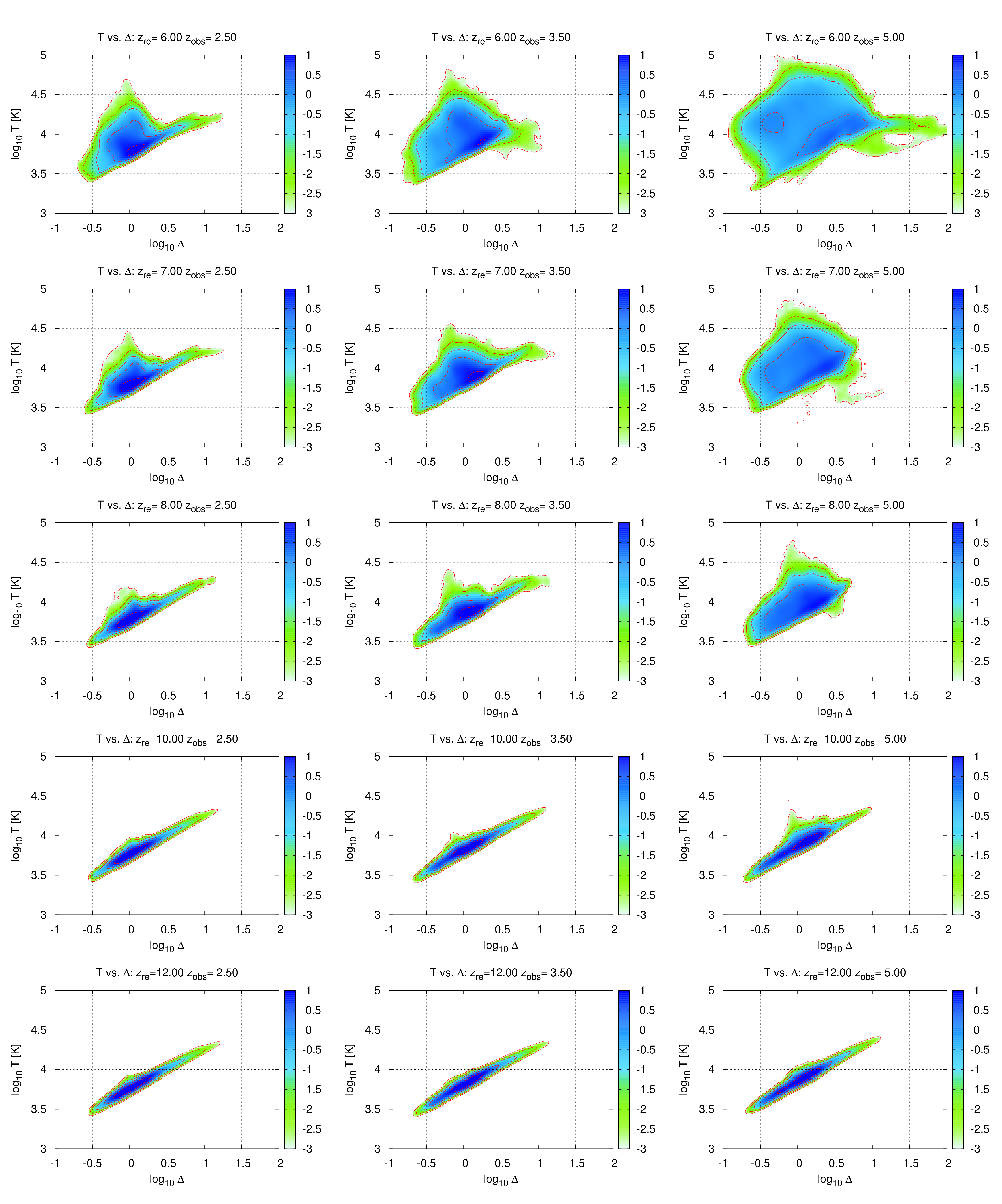}
\caption{\label{fig:post-td}The post-reionization temperature-density relations in the Reference simulation, at several output redshifts $z_{\rm obs}$ (columns) and several reionization redshifts $z_{\rm re}$ (rows). The colour scale and contours are the same as in Figure~\ref{fig:early}.}
\end{figure*}

\begin{figure}
\includegraphics[width=3.2in]{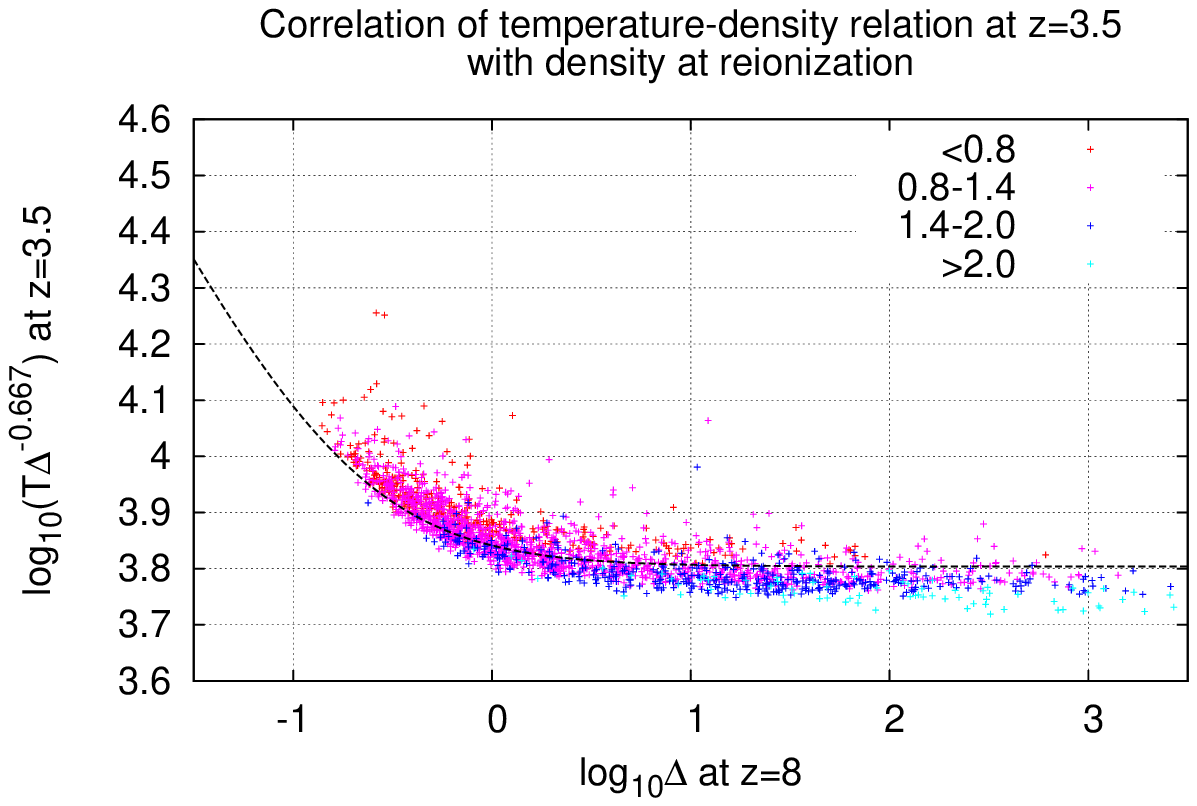}
\includegraphics[width=3.2in]{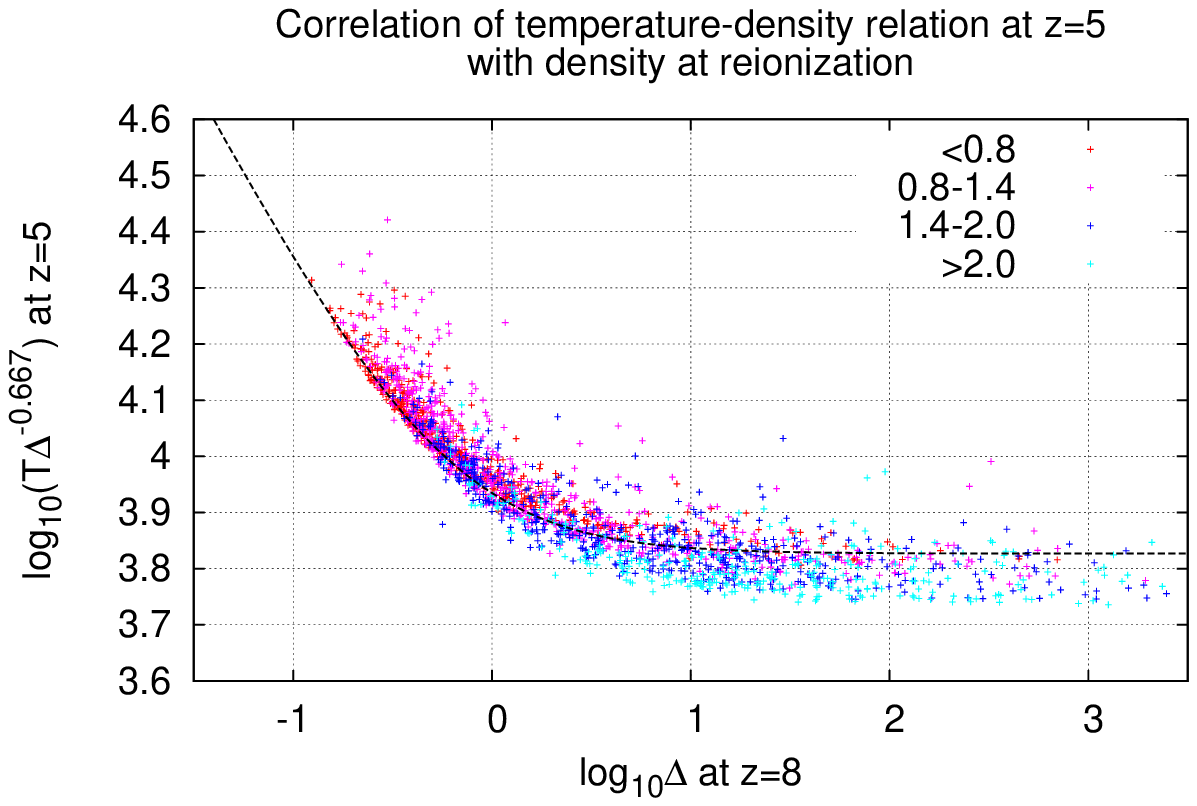}
\caption{\label{fig:scatter1}{\slshape Upper panel}: The temperature-density relation at $z=3.5$ in the variant of the Reference simulation that reionized at $z_{\rm re}=8$. The vertical axis is the entropy, $\log_{10}(T/\Delta^{2/3})$ (temperature in Kelvin), which should be a constant for all gas particles if a temperature-density relation with the slope $\gamma\approx 5/3$ has been reached. The horizontal axis is $\log_{10}\Delta$ at $z=z_{\rm re}=8$. A random subsample of 1/1000 of the gas particles has been plotted. The color scale indicates the baryon density $\Delta$ at $z=3.5$, with red points marking the least-dense gas and light blue marking the densest gas. The dashed black line is our implementation of the \citet{2016MNRAS.456...47M} model. Note the remarkable success of the \citet{2016MNRAS.456...47M} model in describing the simulation results, with the exception of a few points. The higher-density (at $z=3.5$) points lie slightly below the locus at mean density, indicating that the slope of the $T-\Delta$ relation at $z=3.5$ is slightly less than the adiabatic slope of $2/3$. {\slshape Lower panel}: Same at $z=5$.}
\end{figure}

Figure~\ref{fig:post-td} shows the temperature-density relations in the IGM long after reionization. Generally, one expects the temperature-density relation to evolve toward a narrow power-law, $T=T_0\Delta^{\gamma-1}$. A tight power law does indeed form for early reionization ($z_{\rm re}\gtrsim 10$), however for late reionization ($z_{\rm re}\lesssim 8$) we find a significant amount of gas above the median $T-\Delta$ relation -- what we might call a ``high-entropy, mean-density'' (HEMD) phase, the descendent of the high-entropy cloud in Figure~\ref{fig:early}. In hierarchical structure formation models, high-entropy gas is found due to shock heating during infall into filaments or halos, or (if implemented) feedback from star formation or AGNs. Here the HEMD gas was in mini-voids at the time of reionization, and was subsequently compressed (in the sense of {\em comoving} coordinates\footnote{The vast majority of the particles in this simulation are at lower physical density at $z=3.5$ than at $z=8$.}) to near mean density. This is seen most clearly in Figure~\ref{fig:scatter1}, which shows the entropy $\log_{10}(T/\Delta^{2/3})$ at $z=3.5$ and $z=5$ as a function of the overdensity at reionization. As one can see from Figure~\ref{fig:hot-dens}, the compression of gas in mini-voids occurs via propagation of shocks from higher-density regions. But the shocks -- while they provide a major mechanism to compress the gas -- are not the principal source of entropy in the HEMD gas. To see this, we overplot the entropy evolution model of \citet[Eq.~21]{2016MNRAS.456...47M}, which contains Compton and adiabatic cooling and photoionization heating, but no shocks.\footnote{In implementing this model, we make a further simplification: in calculating the integral in Eq.~(21) of \citet{2016MNRAS.456...47M}, we assume that $\Delta(a')=1$ for $a'>a_{\rm re}$, i.e.\ that the gas rapidly relaxes to mean density following reionization.} This model is based on the evolution of the entropy parameter $\eta = T/n^{2/3}\propto Ta^2/\Delta^{2/3}$, and shows that $\eta^{5/3}$ is (i) unaffected by adiabatic expansion, (ii) undergoes a multiplicative suppression due to Compton cooling, and (iii) undergoes an additive term due to photoionization heating. Remarkably, this simple model explains the existence of the HEMD gas and explains the entropy enhancement at the $\sim 10$ per cent level. The upturn at low $\Delta(z=8)$ in this model is due to the {\em initial entropy} of the gas following reionization. The explanation for HEMD gas, then, is that gas in mini-voids is reionized to high entropy. As the IGM subsequently evolves, photoionization heating adds more entropy, and in a mean-density universe this post-reionization entropy injection is dominant. However the initial entropy of mini-voids is so large that it remains significant even if those gas parcels are traced down to $z=3.5$.

The Reference simulation has $\tau_1 = 0.74,\ 0.45,\ 0.27,\ 0.15$ at $z=4.0,\ 3.5,\ 3.0,\ 2.5$; that is, to reproduce the observed transmission in the Lyman-$\alpha$ forest, mean-density gas at $T_4=1$ must have a low optical depth, especially at the lower redshifts. The variations in $\ln\tau_1$ as a function of simulation physics will be explored in more detail next.

\subsection{Variations on the Reference simulation}

We next explore how the aforementioned picture changes as we consider variations on the Reference simulation. All of the variations described herein were run with the same box size and random number generator seed, so that differences from the Reference simulation are entirely the result of the changes described and not due to cosmic variance. For the IGM transparency results in Table~\ref{tab:transparent}, we ran 4 boxes with different random number generator seeds to provide a rough uncertainty estimate.

\subsubsection{Streaming velocities}

The first variation we consider is to turn off the streaming velocities. Physically, one would expect that this allows more small-structure to form, and hence produce a more violent mini-halo evaporation process. This is indeed true. Without the streaming velocity ($v_{\rm bc}=0$), we find that at $z_{\rm re} = 12,\ 10,\ 8,\ 6$, the kinetic energy injection is $\epsilon_{\rm k} = 0.57,\ 0.79,\ 0.99,\ 1.19\, $eV baryon$^{-1}$; the rise time is $t_{\rm rise} = 16,\ 21,\ 31,\ 54$ Myr; and the fall time is $t_{\rm fall} = 102,\ 128,\ 183,\ 332$ Myr. The energy injection $\epsilon_{\rm k}$ is 10 per cent ($z_{\rm re}=6$) to 40 per cent ($z_{\rm re}=12$) greater in the case with no streaming velocities, the rise times are 13--28 per cent faster, and the fall times are 3--12 per cent faster. All of this is expected since the no streaming velocity case does not just lead to more small-scale structure, but to smaller structures with faster sound-crossing times.

For definiteness, let us consider the simulation box that reionizes at $z_{\rm re}=8$ and examine the outputs at $z=3.5$. If we consider all particles with $\Delta<3$ at $z=3.5$ in the Reference simulation, the 10th, 20th, 50th, 80th, and 90th percentile value of $T\Delta^{-2/3}$ are 5937, 6122, 6808, 7963, and 8686 K. In the simulation box with no streaming velocities, these rise to 5943, 6124, 6812, 8078, and 8887 K. Note that the median and lower percentiles changed by at most 0.1 per cent, indicating the robustness of the main temperature-density relation to streaming velocity effects. However, the high-entropy cloud is shifted: the 80th percentile $T/\Delta^{-2/3}$ changed by 1.4 per cent, and the 90th percentile changed by 2.3 per cent. The ``spread'' of 80th versus 50th percentiles changes from 0.068 dex (Reference) to 0.074 dex (no streaming velocities). These changes are subtle, and while they may be of interest for precision cosmology (see \S\ref{sec:implications} below), they are far less dramatic in terms of IGM physics than the effect of the reionization redshift itself. Note also that the relevant effect of streaming velocities on the temperature-density relation comes from the mini-voids, since it is this material that becomes the HEMD gas at low redshift.

In Table~\ref{tab:transparent}, we show the effect of the streaming velocities on the transmission of the IGM at $2.5\le z\le 4.0$. The effect of streaming velocities on the Lyman-$\alpha$ absorption -- parameterized by $\Delta\tau_1$ -- is of the order of 0.5 per cent.

\cchange{
We can try to understand the impact of streaming velocities shown in Table~\ref{tab:transparent} using the analytic model for entropy evolution in Fig.~\ref{fig:scatter1}. This model was based on \citet[Eq.~21]{2016MNRAS.456...47M}, with the approximation $[\Delta(a')]^{-1/9}\rightarrow 1$ in their integral for $a_{\rm re}<a'<a$, and it quantitatively explains the high entropy of gas that was in voids at the time of reionization. If we consider the case of $z_{\rm re}=8$, $T_{\rm re}=2\times 10^4$ K, and observations at $z=3$, this model predicts a final entropy
\begin{equation}
\eta_{\rm f}^{5/3} = 6.08\times 10^{10} [\Delta(z_{\rm re})]^{-10/9} + 5.86\times 10^{11}\,\,{\rm K}^{5/3}\,{\rm cm}^{-10/3},
\label{eq:EF}
\end{equation}
where $\eta = T/n_{\rm H}^{2/3}$. If this manifested itself as simply a shifting of the final gas temperature, then we would have $T_{\rm f}\propto\eta_{\rm f}$, and then the optical depth would be re-scaled by $\tau\propto T_{\rm f}^{-0.7}\propto (\eta_{\rm f}^{5/3})^{-0.42}$, where the exponent $-0.7$ comes from the H{\sc\,i} recombination coefficient. Taking the first term in Eq.~(\ref{eq:EF}) to be smaller than the second, we would infer that
\begin{eqnarray}
\Delta\ln\tau_1 \!\!\!\! &\approx&\!\!\!\! -0.42\frac{{\rm Change\,}[\eta_{\rm f}^{5/3}]}{\eta_{\rm f}^{5/3}}
\nonumber \\
&\approx&\!\!\!\! 0.42 \times \frac{6.08\times 10^{10}}{5.86\times 10^{11}}  {\rm Change\,}\langle [\Delta(z_{\rm re})]^{-10/9} \rangle,
\label{eq:DTEF}
\end{eqnarray}
where the $-$ sign in the first line arises since $\tau_1$ is defined as the change in optical depth scale required to {\em compensate} for changes in the simulation physics and leave $\bar F$ unchanged. (Here we write ``change'' to avoid confusion with the baryon overdensity $\Delta$ on the right-hand side.)

The Phase I simulations with $v_{\rm bc}$ on (``Reference'') and off (``nov'') have $\langle [\Delta(z_{\rm re})]^{-10/9} \rangle = 1.067$ and 1.079, respectively (particle-weighted) or 2.41 and 2.76 (volume-weighted at $z=z_{\rm re}=8$). These differences and Eq.~(\ref{eq:DTEF}) imply $\Delta\ln\tau_1 = 0.0005$ (particle-weighted) or 0.015 (volume-weighted), as compared with the value of $0.0047\pm 0.0004$ based on extracted Lyman-$\alpha$ cubes (see Table~\ref{tab:transparent}). The effect predicted by Eq.~(\ref{eq:DTEF}) has the correct sign, however the order of magnitude depends on how the averaging is done (weighted by particles, i.e.\ baryonic mass, or by volume). In principle the particle weighting makes more sense, because mass elements or particles are conserved as small-scale structure is smeared out whereas volume elements are not. However, it is also true that by particle weighting $\langle [\Delta(z_{\rm re})]^{-1} \rangle=1$ (by conservation of {\em total} comoving volume); the particle-weighted $\langle [\Delta(z_{\rm re})]^{-10/9} \rangle$ thus involves a near-cancellation of voids (which lead to higher entropy gas) and clumps (which lead to lower entropy gas), and can differ from 1 only because the exponent $-10/9$ differs slightly from $-1$. The fact that the $\Delta\ln\tau_1$ based on extracted Lyman-$\alpha$ cubes (0.0047) is smaller than the volume-weighted prediction (0.015) suggests that some of this cancellation is indeed realized in the full simulation, however the fact that it is larger than the particle-weighted prediction (0.0005) suggests that the cancellation is not as good as the simple analytic approximation (Eq.~\ref{eq:DTEF}) would imply. We conclude that while analytic estimates can inform the likely range of orders of magnitude of $\Delta\ln\tau_1$, quantitative predictions for Lyman-$\alpha$ forest statistics require fully evolved simulations.
}

\begin{table*}
\caption{\label{tab:transparent}Transparency variations of the IGM in the Phase I simulations, parameterized by $\tau_1$ (the optical depth of a patch of gas at $T_4=\Delta=1$ and no peculiar velocities needed to produce the correct mean transmitted flux). Reionization redshifts are indicated in brackets. The changes are defined by $\Delta\ln\tau_1 = \ln[\tau_1({\rm B})/\tau_1({\rm A})]$. That is, $\Delta\ln\tau_1$ is positive if Simulation B is more transparent than Simulation A (i.e.\ the stated change in physics increases Lyman-$\alpha$ forest transmission), and $\Delta\ln\tau_1$ is negative if Simulation A is more transparent than Simulation B. Errors are the $1\sigma$ error on the mean from 4 simulations with different random number seeds, but note that 4 simulations are not sufficient to derive a robust error estimate. Variations of the Reference simulation with different Lyman-$\alpha$ cube construction parameters ($N_{\rm cell}$ and $\Delta_{{\rm th,Ly}\alpha}$) are also shown.}
\begin{tabular}{ccc|rrrr}
\hline\hline
Physics & Simulation A & Simulation B & \multicolumn4c{$10^5\times \Delta\ln \tau_1$ at:} \\
& & & $z=4.0$ & $z=3.5$ & $z=3.0$ & $z=2.5$ \\
\hline
Turn on X-ray pre-heating & Reference [8] & pre [8] & $-886\pm166$ & $-1148\pm\wsp61$ & $-997\pm\wsp10$ & $-779\pm\wsp33$ \\
Density-dependent $T_{\rm re}$ & Ref [8] & I5 [8] & $349\pm167$ & $28\pm\wsp44$ & $-56\pm\wsp\wsp5$ & $-68\pm\wsp11$ \\
\hline
Turn off streaming velocities & Reference [8] & nov [8] & $536\pm111$ & $557\pm\wsp74$ & $466\pm\wsp38$ & $364\pm\wsp15$ \\
Turn off streaming velocities (X-ray on) & pre [8] & pre-nov [8] & $285\pm\wsp56$ & $331\pm\wsp40$ & $294\pm\wsp19$ & $239\pm\wsp\wsp5$ \\
Turn off streaming velocities (low $\Delta_{\rm th}$) & D100 [8] & D100-nov [8] & $512\pm\wsp58$ & $523\pm\wsp37$ & $439\pm\wsp14$ & $351\pm\wsp14$ \\
Turn off streaming velocities (slow heating) & Soft [8] & Soft-nov [8] & $523\pm\wsp93$ & $538\pm\wsp66$ & $453\pm\wsp33$ & $355\pm\wsp\wsp9$ \\
Turn off streaming velocities (alt. $T_{\rm re}$) & I5 [8] & I5-nov [8] & $439\pm\wsp79$ & $491\pm\wsp53$ & $432\pm\wsp23$ & $347\pm\wsp10$ \\ 
Turn off streaming velocities (low res.) & LoRes1 [8] & LoRes1-nov [8] & $561\pm\wsp90$ & $548\pm\wsp65$ & $444\pm\wsp34$ & $341\pm\wsp14$ \\
Turn off streaming velocities (low res.) & LoRes2 [8] & LoRes2-nov [8] & $584\pm\wsp55$ & $532\pm\wsp64$ & $423\pm\wsp48$ & $330\pm\wsp21$ \\
Turn off streaming velocities ($N_{\rm cell}=64$) & Reference [8] & nov [8] & $578\pm116$ & $576\pm\wsp77$ & $474\pm\wsp39$ & $366\pm\wsp15$ \\
\!\!\!Turn off streaming velocities ($\Delta_{{\rm th,Ly}\alpha}=50$)\!\!\! & Reference [8] & nov [8] & $536\pm111$ & $557\pm\wsp74$ & $466\pm\wsp38$ & $364\pm\wsp15$ \\
Turn off streaming velocities ($\nu_0=0$) & Reference* [8] & nov* [8] & $600\pm110$ & $587\pm\wsp72$ & $479\pm\wsp31$ & $375\pm\wsp10$ \\
\hline
Earlier reionization & Reference [8] & Reference [9] & \!$-5058\pm118$ & \!$-3497\pm\wsp68$ & \!$-2417\pm\wsp50$ & \!$-1578\pm\wsp64$ \\
Earlier reionization (X-ray on) & pre [8] & pre [9] & \!$-4540\pm182$ & \!$-2846\pm\wsp49$ & \!$-1861\pm\wsp41$ & \!$-1152\pm\wsp64$ \\
Earlier reionization (low $\Delta_{\rm th}$) & D100 [8] & D100 [9] & \!$-5179\pm303$ & \!$-3507\pm110$ & \!$-2439\pm112$ & \!$-1609\pm149$ \\
Earlier reionization (slow heating) & Soft [8] & Soft [9] & \!$-5023\pm121$ & \!$-3437\pm\wsp61$ & \!$-2364\pm\wsp49$ & \!$-1542\pm\wsp60$ \\
Earlier reionization (low resolution) & LoRes1 [8] & LoRes1 [9] & \!$-4949\pm\wsp93$ & \!$-3434\pm\wsp74$ & \!$-2378\pm\wsp53$ & \!$-1554\pm\wsp60$ \\
Earlier reionization (low resolution) & LoRes2 [8] & LoRes2 [9] & \!$-4730\pm\wsp74$ & \!$-3297\pm\wsp25$ & \!$-2295\pm\wsp47$ & \!$-1509\pm\wsp70$ \\
Earlier reionization (alt. $T_{\rm re}$) & I5 [8] & I5 [9] & \!$-5259\pm244$ & \!$-3539\pm\wsp68$ & \!$-2431\pm\wsp57$ & \!$-1588\pm\wsp85$ \\
Earlier reionization ($N_{\rm cell}=64$) & Reference [8] & Reference [9] & \!$-5262\pm119$ & \!$-3586\pm\wsp69$ & \!$-2453\pm\wsp50$ & \!$-1591\pm\wsp64$ \\
Earlier reionization ($\Delta_{{\rm th,Ly}\alpha}=50$) & Reference [8] & Reference [9] & \!$-5058\pm118$ & \!$-3497\pm\wsp68$ & \!$-2417\pm\wsp50$ & \!$-1578\pm\wsp64$ \\
Earlier reionization ($\nu_0=0$) & Reference* [8] & Reference* [9] & \!$-5324\pm115$ & \!$-3609\pm\wsp71$ & \!$-2462\pm\wsp52$ & \!$-1591\pm\wsp59$ \\
\hline
Later reionization & Reference [8] & Reference [7] & $8201\pm373$ & $5708\pm120$ & $4247\pm106$ & $3224\pm\wsp51$ \\
Later reionization (X-ray on) & pre [8] & pre [7] & $7915\pm532$ & $4902\pm\wsp82$ & $3324\pm\wsp54$ & $2408\pm\wsp66$ \\
Later reionization (low $\Delta_{\rm th}$) & D100 [8] & D100 [7] & $8277\pm534$ & $5638\pm\wsp70$ & $4108\pm\wsp87$ & $3103\pm\wsp61$ \\
Later reionization (slow heating) & Soft [8] & Soft [7] & $8063\pm389$ & $5543\pm142$ & $4151\pm\wsp92$ & $3147\pm\wsp56$ \\
Later reionization (low resolution) & LoRes1 [8] & LoRes1 [7] & $8224\pm381$ & $5710\pm121$ & $4237\pm\wsp99$ & $3209\pm\wsp50$ \\
Later reionization (low resolution) & LoRes2 [8] & LoRes2 [7] & $8137\pm338$ & $5603\pm122$ & $4157\pm\wsp93$ & $3147\pm\wsp49$ \\
Later reionization (alt. $T_{\rm re}$) & I5 [8] & I5 [7] & $8191\pm525$ & $5555\pm\wsp93$ & $4080\pm\wsp91$ & $3093\pm\wsp56$ \\
Later reionization ($N_{\rm cell}=64$) & Reference [8] & Reference [7] & $8596\pm384$ & $5882\pm120$ & $4318\pm105$ & $3250\pm\wsp51$ \\
Later reionization ($\Delta_{{\rm th,Ly}\alpha}=50$) & Reference [8] & Reference [7] & $8202\pm375$ & $5708\pm119$ & $4247\pm106$ & $3224\pm\wsp51$ \\
Later reionization ($\nu_0=0$) & Reference* [8] & Reference* [7] & $8727\pm391$ & $5941\pm118$ & $4347\pm100$ & $3267\pm\wsp51$ \\
\hline
Reduce self-shielding threshold $\Delta_{\rm th}$ & Reference [8] & D100 [8] & $204\pm325$ & $26\pm196$ & $15\pm158$ & $29\pm135$ \\
Slow heating of dense gas & Reference [8] & Soft [8] & $32\pm\wsp49$ & $-59\pm\wsp15$ & $-82\pm\wsp15$ & $-67\pm\wsp14$ \\
\hline
Lower resolution & Reference [8] & LoRes1 [8] & $-199\pm\wsp42$ & $-42\pm\wsp\wsp8$ & $40\pm\wsp\wsp8$ & $66\pm\wsp13$ \\
Lower resolution & Reference [8] & LoRes2 [8] & $-715\pm\wsp76$ & $-208\pm\wsp36$ & $46\pm\wsp65$ & $153\pm\wsp64$ \\
\hline\hline
\end{tabular}
\end{table*}

\subsubsection{X-ray pre-heating}

A second variation is the inclusion of X-ray pre-heating. As described in \S\ref{ss:X-ray}, we implement X-ray heating as a gradually increasing energy input over the whole box, with uniform energy deposition per baryon. The model considered here is sufficient to heat the gas to the CMB temperature -- i.e.\ to flip the H{\,\sc i} 21 cm line from absorption to emission -- at $z=16$. The median gas temperature in the neutral phase then rises to 128 K at $z=12$, 455 K at $z=8$, and 1098 K at $z=6$. The X-ray energy input is significantly greater than predicted by recent models of X-ray binary heating \citep{2014Natur.506..197F}, which do not heat to the CMB temperature until $z=12$. However the model uncertainties are large and so we chose to run both a case with {\em no} X-ray heating (the default) and one with more heating than recent estimates.

In the model with X-ray heating (the ``pre'' simulation), at $z_{\rm re} = 12,\ 10,\ 8,\ 6$, the kinetic energy injection is $\epsilon_{\rm k} = 0.32,\ 0.44,\ 0.53,\ 0.62\, $eV baryon$^{-1}$; the rise time is $t_{\rm rise} = 30,\ 38,\ 59,\ 117$ Myr; and the fall time is $t_{\rm fall} = 137,\ 172,\ 264,\ 465$ Myr. Generally, the X-ray heating has a similar effect as the streaming velocities: by suppressing small-scale structure, it reduces the amount of kinetic energy injection from disruption of mini-halos, and since the structures that survive are larger the timescales are longer. The effect of mini-halo evaporation is still there in the models with X-ray heating, but it is evident that it can be significantly suppressed: the energy injection is reduced by up to 42 per cent, the rise times are up to a factor of 2.2 longer, and the fall times are up to a factor of 1.4 longer, with the most dramatic effects happening at $z_{\rm re}=6$.

We may also consider the effect of X-ray pre-heating on the high-entropy, mean-density gas at lower redshifts. As before, we examine the $\Delta<3$ gas in the simulation that reionized at $z_{\rm re}=8$. The 10th, 20th, 50th, 80th, and 90th percentile value of $T\Delta^{-2/3}$ are 5893, 6075, 6717, 7771, and 8422 K. That is, the ratio of the 80th percentile to median entropy is 1.16 (versus 1.17 in the Reference simulation), and the ratio of 90th percentile to mean entropy is 1.25 (versus 1.28 in the Reference simulation). The differences are larger at higher $z$ (e.g.\ at $z=5.5$, the 80th:50th percentile ratio is 1.35 in the ``pre'' simulation and 1.44 in the ``Reference'' simulation) and smaller at lower $z$. Thus the existence and qualitative properties of the HEMD gas appear to be robust to X-ray pre-heating.

\cchange{
The modest impact of the X-ray heating may at first seem surprising, given that it dramatically increases the temperature and hence the Jeans mass of the gas. However, one must be mindful both of what happens to gas in mini-haloes and in voids. Mini-halo gas has already been shock-heated to well above the ambient IGM temperature, and so the X-ray heating (which is roughly uniform in injected energy per baryon and hence in temperature increase) has less of an impact here than in mean-density regions. For example, at $z=8$, the median gas temperature in our model is 455 K; this is equal to the virial temperature for a $3.6\times 10^5\,M_\odot$ halo.

In the case of gas in early voids, the X-ray heating leads to a dramatic increase in gas temperature, relative to the unheated mean-density temperature of 2 K (at $z_{\rm re}=8$). However, the spatial distribution of gas does not change instantaneously in response to heating -- it is more closely related to the filtering scale, which depends on the full thermal history of the gas \citep{1998MNRAS.296...44G}. In a matter-dominated universe, the filtering scale can be written as
\begin{equation}
[k_{\rm F}(a)]^{-2} = \frac{9t^2}{2a^2} \int_0^1 \psi(1-\sqrt\psi) \,\frac{\gamma k_{\rm B}T(\psi a)}{\mu}\,d\psi,
\label{eq:KFA}
\end{equation}
where $\mu$ is the mean molecular weight, $\gamma$ is the pressure-density relation slope, and $T(\psi a)$ is the temperature of the gas at $a'=\psi a$; the integral runs from the Big Bang ($\psi=0$) to the epoch $a$ at which the filtering scale is measured ($\psi=1$). For the isothermal case ($\gamma=1$) and the {\sc HyRec} temperature history (with no X-ray heating), the filtering scale at $z=8$ is $k_{\rm F}^{-1}=1.8$ ckpc. If we include X-ray heating as defined in the model of this paper, so that $T$ is increased by $300 a_{-1}^{3.5}$ K, then $k_{\rm F}^{-1} = 3.1$ ckpc. The change is modest because (i) most of the temperature increase occurs just before reionization (i.e.\ over a small range in $\psi$ near 1); (ii) the integrand in Eq.~(\ref{eq:KFA}) down-weights $\psi\approx 1$ (see the factor of $1-\sqrt\psi$); and (iii) even in the no-heating case, the gas was hotter at early times and this leads to a substantial contribution to the filtering length integral. The result is that even though X-ray heating increases the final gas temperature by a factor of 200, it only increases the filtering length by a factor of 1.8.
}

One might wonder how the X-ray heating interacts with the streaming velocities, in particular whether X-ray heating renders the streaming velocities not relevant by wiping out any structures small enough to be sensitive to streaming velocities. To study this, we ran a simulation with the X-ray heating on and the streaming velocities off (``pre-nov'') and compared it to the simulation with X-rays on and streaming velocities on (``pre''). As seen in Table~\ref{tab:transparent}, the effect of streaming velocities on the Lyman-$\alpha$ forest absorption is indeed reduced by the X-ray heating. However this suppression is modest, e.g.\ at $z=2.5$ their 0.36 per cent effect (streaming velocities on vs.\ off with no X-rays) is reduced to a 0.24 per cent effect (streaming velocities on vs.\ off with X-rays).

\subsubsection{Varying the treatment of high-density gas}

Because the simulations in this paper are only hydrodynamic, i.e.\ with no radiative transfer, the treatment of dense gas is a potential source of uncertainty. Missing from our treatment is the delayed ionization of the densest gas, the hardening of the radiation field as it approaches a self-shielded region, and the proper dynamics of gas acceleration and gas heating at the D-type ionization fronts that ultimately destroy the self-shielded clumps. At the level of this paper, we only attempt to consider a range of prescriptions to assess whether the treatment of high-density gas is important. In particular, we want to know whether the sensitivity to streaming velocities or the reionization redshift changes with the treatment of the high-density gas. As one can see from Table~\ref{tab:transparent}, the effects of the high-density cutoff (changing $\Delta_{\rm th}$) or the slow-heating (``Soft'' runs) are small. The sensitivity to $v_{\rm bc}$ and to $z_{\rm re}$ changes by at most 6 per cent for the cases in the table.

\subsubsection{Varying the reionization temperature model}

The reference model assumes reionization to a temperature of $2\times 10^4$ K. This parameter normally has a very small impact on the low redshift IGM temperature due to convergent thermal evolution \citep[e.g.][]{2016MNRAS.456...47M}, but since in this investigation we are interested in the memory of reionization it is important to check other models for the reionization temperature. The alternative model considered here is a physical model of ionization fronts described in Appendix~\ref{app:I-front}. This model has two ingredients: the speed of the ionization front $v_{\rm i}$, and the spectrum of incident radiation at the ionization front, parameterized as a blackbody of temperature $T_{\rm bb}$. If the flux and spectrum of the ionizing radiation were uniform over the whole box, then we would have $v_{\rm i} = \bar v_{\rm i}/(1+\delta_{\rm b})$, where $\bar v_{\rm i}$ is the ionization front velocity at mean density. This model is still imperfect, as it does not account for the tilting of ionization fronts ($\cos\theta$ effect) and hardening of the radiation spectrum as one approaches an over-dense clump, nor can it account for variations in ionizing radiation flux or spectrum on scales larger than the box size. It does predict that the dense regions reionize to lower temperatures, because they experience more Lyman-$\alpha$ cooling during the passage of the ionization front.

The parameters chosen for the alternative (``I5'') model are an incident blackbody temperature of $T_{\rm bb} = 5\times 10^4$ K and an ionization front speed at mean density of $\bar v_{\rm i} = 5\times 10^8$ cm$\,$s$^{-1}$. At $z=8$, this corresponds to 5 cMpc per $\Delta z=1$, or to an ionizing flux of $F = 5\times 10^4$ photons$\,$cm$^{-2}\,$s$^{-1}$. This is plausible given the size and lifetimes of reionization bubbles that are seen in simulations, although $F$ undoubtedly has large variations. For comparison, the \citet{2012ApJ...746..125H} background model gives a flux of $F = \int [4\pi I_\nu/(h\nu)]\,d\nu/Q_{\rm HII} = 4\times 10^4$ photons$\,$cm$^{-2}\,$s$^{-1}$ at $z=8$.

\subsubsection{The initial decaying mode}

Our default simulation contained an initial decaying mode ($\nu_0=0.21$), which is an approximate treatment of the more complicated conditions at decoupling in the real Universe. As a test for how much this matters, we turned this initial decaying mode off ($\nu_0=0$). As seen in Table~\ref{tab:transparent}, the effects of streaming velocities and the reionization redshift change by at most 12 per cent.

\subsubsection{Resolution}

Finally, we investigate the convergence with respect to resolution by running cases with particles at $(4/3)^3$ and $2^3$ times worse mass than the Reference simulation. As seen in Table~\ref{tab:transparent}, the sensitivity to streaming velocities (i.e.\ $\Delta\ln\tau_1$ between the streaming velocities on and off cases) changes by at most 10 per cent at the Reference resolution versus the $2^3\times$ worse resolution. Therefore the Phase II simulations were run at the worse resolution, in order to explore the largest possible volume.

\subsection{Phase II simulations (box size convergence)}
\label{ss:phase2}

The Phase II simulations were run with the same resolution as LoRes2, i.e.\ with dark matter particles of 9720 $M_\odot$ and baryonic particles of mass 1810 $M_\odot$. Eight simulation boxes were run of each size (see Table~\ref{tab:sim-param}). The smallest simulation boxes (II-A and II-An) are equivalent to the Phase I LoRes2 and LoRes2-nov runs, but a suite of eight boxes with new random number generator seeds was used anyway for consistency with the rest of Phase II.

For Phase II, all boxes of the same size use the same initial conditions, but a new set of seeds is chosen for each box size. The boxes II-A, II-B, II-C, II-D, and II-F represent a progressive increase in the box size, from II-A (the same size as Phase I) through II-F ($6^3=216$ times more volume).\footnote{The progression suggests a ``II-E'' box size, of length 2126 ckpc, but this was not included as it would have required re-sampling of the initial conditions.} The box size can also be described in terms of the missing variance $\sigma^2(M)$, where $M$ is the total mass of the box; this describes the amount of power that is present in the real Universe on scales larger than the box size, but which is removed because we force the box to mean density. \cchange{In problems where one is concerned with the effects of large-scale modes, the missing variance is often the more relevant parameter than the box size. Note that the missing variance converges slowly with box size, due to the roll-over in the matter power spectrum at the relevant scales.} At $z=2.5$, we have $\sigma^2(M)=2.67$, 1.74, 1.32, 1.07, and 0.77 for boxes II-A, II-B, II-C, II-D, and II-F respectively.

Results for the streaming velocity and reionization redshift dependences are shown in Table~\ref{tab:phase2}. The transparency change due to streaming velocities converges rapidly: at the $2\sigma$ level, there is no difference between the II-B and II-F results, despite a factor of 27 increase in simulation volume. For the dependence on the redshift of reionization, there are still substantial changes over the range of box sizes considered; at $z\sim 2.5$, the change in going from the II-B to II-F box size is 18\%\ (for the $z_{\rm re}=8\rightarrow 7$ case) and 38\%\ (for the $z_{\rm re}=8\rightarrow 9$ case). Further increases in the box size would be necessary to achieve full convergence on the reionization redshift dependence, $\partial(\ln\tau_1)/\partial z_{\rm re}$. This dependence is not needed for the main result of this paper, and hence the larger box size simulations are left to future work.

\section{BAO peak shift}
\label{sec:implications}

Our final step is to convert our results for the effect of streaming velocities on the Lyman-$\alpha$ forest transmission, i.e.\ $\Delta\tau_1$, into a prediction for changes in the BAO scale. This requires us to return to large-scale structure biasing theory, and estimate the relevant bias coefficients from large-scale structure simulations, observations, and the small-box simulations in this paper that investigate streaming velocities.

\subsection{The streaming velocity bias coefficient}

\begin{table*}
\caption{\label{tab:phase2}The Phase II simulation results. The table shows variations in transparency $\Delta \ln\tau_1$ for Simulation B relative to Simulation A. Errors shown are standard deviations on the mean of 8 simulations. The table is grouped into three types of changes (turn off streaming velocities; earlier reionization; and later reionization). Within each group, successive lines indicate larger boxes at fixed resolution. The ``$\sigma^2(M)$'' column denotes the linear variance at $z=2.5$ computed for the total mass in the simulation volume, and is indicative of the amount of missing large-scale power.}
\begin{tabular}{ccccrrrr}
\hline\hline
Box size & $\sigma^2(M)$ & Sim. A & Sim. B & \multicolumn4c{$10^5\times \Delta\ln \tau_1$ at:}  \\
{[ckpc]} & $z=2.5$ & & & $z=4.0$ & $z=3.5$ & $z=3.0$ & $z=2.5$ \\
\hline
\multicolumn7c{Turn off streaming velocities} \\
\wsp425 & 2.67 & II-A[8] & II-An[8] & $596\pm\wsp34$ & $531\pm\wsp29$ & $417\pm\wsp14$ & $312\pm\wsp\wsp8$ \\
\wsp850 & 1.74 & II-B[8] & II-Bn[8] & $174\pm\wsp53$ & $232\pm\wsp42$ & $231\pm\wsp34$ & $221\pm\wsp38$ \\
1275 & 1.32 & II-C[8] & II-Cn[8] & $165\pm\wsp81$ & $188\pm\wsp70$ & $198\pm\wsp67$ & $221\pm\wsp64$ \\
1701 & 1.07 & II-D[8] & II-Dn[8] & $212\pm\wsp84$ & $224\pm\wsp73$ & $241\pm\wsp64$ & $276\pm\wsp53$ \\
2551 & 0.77 & II-F[8] & II-Fn[8] & $185\pm\wsp28$ & $225\pm\wsp28$ & $258\pm\wsp31$ & $300\pm\wsp30$ \\
\hline
\multicolumn7c{Earlier reionization} \\
\wsp425 & 2.67 & II-A[8] & II-A[9] & $-4983\pm152$ & $-3408\pm\wsp40$ & $-2375\pm\wsp23$ & $-1574\pm\wsp30$ \\
\wsp850 & 1.74 & II-B[8] & II-B[9] & $-4290\pm176$ & $-2535\pm232$ & $-1397\pm226$ & $-810\pm162$ \\
1275 & 1.32 & II-C[8] & II-C[9] & $-4223\pm168$ & $-2047\pm230$ & $-761\pm241$ & $-296\pm196$ \\
1701 & 1.07 & II-D[8] & II-D[9] & $-4300\pm\wsp85$ & $-2020\pm\wsp83$ & $-742\pm\wsp76$ & $-379\pm\wsp76$ \\
2551 & 0.77 & II-F[8] & II-F[9] & $-4625\pm\wsp93$ & $-2376\pm\wsp91$ & $-1035\pm\wsp85$ & $-505\pm\wsp70$ \\
\hline
\multicolumn7c{Later reionization} \\
\wsp425 & 2.67 & II-A[8] & II-A[7] & $7714\pm367$ & $5674\pm109$ & $4153\pm\wsp42$ & $3210\pm\wsp58$ \\
\wsp850 & 1.74 & II-B[8] & II-B[7] & $8306\pm107$ & $5711\pm124$ & $3808\pm133$ & $2713\pm107$ \\
1275 & 1.32 & II-C[8] & II-C[7] & $8371\pm\wsp93$ & $5400\pm189$ & $3224\pm200$ & $2179\pm180$ \\
1701 & 1.07 & II-D[8] & II-D[7] & $8415\pm151$ & $5198\pm111$ & $3069\pm\wsp89$ & $2030\pm\wsp82$ \\
2551 & 0.77 & II-F[8] & II-F[7] & $8788\pm120$ & $5581\pm109$ & $3365\pm119$ & $2223\pm\wsp81$ \\
\hline\hline
\end{tabular}
\end{table*}

On linear scales, in the absence of complicating effects from reionization, small-scale structure, or ionizing background fluctuations, the 3D power spectrum of the Lyman-$\alpha$ forest is expected to follow the form:
\begin{equation}
P_F(k,\mu) = b_F^2(1+\beta_F\mu^2)^2 P_{\rm m}(k),
\label{eq:Linear}
\end{equation}
where $\mu$ is the cosine of the angle between the Fourier wave vector and the line of sight, $P_{\rm m}(k)$ is the matter power spectrum, and $b_F$ and $\beta_F$ are the linear bias coefficients. Deviations from this formula occur at small scales; for example they are expected to reach a factor of $\sim 1.5$ at $k=1.3\,$cMpc$^{-1}$ and $z=2.6$ \citep[Fig.~21]{2015JCAP...12..017A}. These scales are important for broadband Lyman-$\alpha$ absorption studies, but are generally at smaller scales than those of interest for BAO.

One commonly defines $b_{F\Gamma} = \partial \ln\bar F/\partial\ln\tau_1$ as the variation of the (log) transmitted flux with respect to a uniform rescaling of the optical depth \citep[Eq.~2.7]{2015JCAP...12..017A}. This quantity is not directly associated with the mapping between the matter and flux power spectra. It is, however, the bias parameter that describes how large-scale ionizing background fluctuations affect the Lyman-$\alpha$ forest, since in photoionization equilibrium the optical depth is inversely proportional to the ionizing background. It is also of theoretical interest in analytic theories of Lyman-$\alpha$ forest biasing \citep[e.g.][]{2012JCAP...03..004S}. Here, it is needed since we scaled the simulation results to reproduce the correct mean transmitted flux and reported changes in $\ln\tau_1$.

The simulations by \citet{2015JCAP...12..017A}\footnote{We used the $D_1$ nonlinear fitting function with $q_2=0$. The relation between their $b_{\tau\delta}$ and our $b_F$ is $b_F = (\ln\bar F) b_{\tau\delta}$.} find $b_F = -0.201$ and $\beta_F=1.205$ ($z=3.0$), and $b_F = -0.125$ and $\beta_F=1.364$ ($z=2.5$, interpolated). Numerical values are not reported for $b_{F\Gamma}$, but by applying the integral form to the observed flux probability density functions \citep[Table~A3]{2007MNRAS.382.1657K}, we find $b_{F\Gamma} = 0.156$ ($z=3.0$) and 0.122 ($z=2.5$).

The key parameter is the sensitivity of the Lyman-$\alpha$ transmitted flux to the streaming velocity of baryons relative to dark matter. The effect is parameterized by $b_{Fv}$, which is the fractional change in transmitted flux due to the streaming velocity; this can be expressed as
\begin{equation}
b_{Fv} = \ln \bar F({\rm with}\,v_{bc}) - \ln \bar F({\rm no}\,v_{bc})
= -b_{F\Gamma} \Delta\ln\tau_1,
\label{eq:bFv}
\end{equation}
where $\Delta\ln\tau_1$ is the change in IGM transparency when the streaming velocities are turned {\em off}.

\subsection{Estimation of the peak shift}

The streaming velocity bias translates into a fractional change $\Delta\alpha$ in the BAO scale via some sensitivity coefficient $\partial\alpha/\partial (b_{Fv}/b_F)$. This sensitivity coefficient was estimated in \citet{2016PhRvL.116l1303B}, but the result varies somewhat depending on the redshift $z$ and the bias-weighted sampling density $b^2n$. Note that the sampling density does not enter into the power spectrum of the galaxies, but it affects the optimal weighting of different $k$ modes and so has an impact on the shift $\alpha$ derived from a model fit. In the case of the Lyman-$\alpha$ forest, there is not a 3D number density $n$, but there is an effective noise level that replaces galaxy Poisson noise based on the 2D density of sight-lines $n_{\rm eff}^{\rm 2D}$, the 1D Lyman-$\alpha$ forest power spectrum (describing aliased signal) $P_{\rm los}(k_\parallel)$, and a down-weighting factor $\nu_n$ based on the signal-to-noise of the spectra: $n_{\rm eff}^{\rm 3D} = n_{\rm eff}^{\rm 2D} \nu_n/P_{\rm los}(k_\parallel)$ \citep{2011MNRAS.415.2257M}.

The sight-line densities relevant to DESI \citep{2016arXiv161100036D} can be estimated by integrating the stated redshift distribution of quasars over the range useful for the Lyman-$\alpha$ forest (985--1200 \AA\ in the published forecasts); this leads to a density of $7.5\times 10^{-4}$ Mpc$^{-2}$ ($z=3.0$) and $1.9\times 10^{-3}$ Mpc$^{-2}$ ($z=2.5$). Using the $P_{\rm los}(k_\parallel)$ and $b$ estimates in \citet{2011MNRAS.415.2257M}, and the noise weighting $\nu_n$ relevant for spectra of signal-to-noise ratio of 2 per 1\,\AA\ synthetic pixel, we forecast $b^2n_{\rm eff}^{\rm 3D} = 7.4\times 10^{-5}$ Mpc$^{-3}$ ($z=3.0$) and $5.6\times 10^{-5}$ Mpc$^{-3}$ ($z=2.5$). The scripts in \citet{2016PhRvL.116l1303B} then predict sensitivity coefficients of $\partial\alpha/\partial (b_{Fv}/b_F) = 0.41$ ($z=3.0$) and 0.41 ($z=2.5$), using two-sided derivatives with respect to $b_{Fv}/b_F$ with a step size of $\pm 0.002$.\footnote{I wish to thank Jonathan Blazek for re-running this set of scripts on the grid of values needed for this project.}$^,$\footnote{The appearance of 0.41 twice is not a typo.} This calculation of $n_{\rm eff}^{\rm 3D}$ is obviously very rough, and the actual weighting of $k$-modes used by future experiments such as DESI may be different. However, we find that even for $\pm 1$ dex changes in $n_{\rm eff}^{\rm 3D}$, the coefficient $\partial\alpha/\partial (b_{Fv}/b_F)$ only varies over the range 0.31--0.46. Given the substantial astrophysical uncertainties in this calculation (see \S\ref{ss:uncert}), we believe this highly simplified treatment of the $k$-dependent weighting is appropriate.

Overall, we may then write
\begin{equation}
\Delta\alpha \approx 0.41 \frac{b_{Fv}}{b_F} = -0.41 \frac{b_{F\Gamma}}{b_F} \Delta\ln\tau_1.
\label{eq:stream}
\end{equation}
The coefficient $\Delta\alpha/\Delta\ln\tau_1$ is 0.32 ($z=3.0$) or 0.40 ($z=2.5$).

Based on Eq.~(\ref{eq:stream}), the Phase II simulation results in Table~\ref{tab:phase2} then imply a BAO peak shift of $0.13$\%\ ($z=3.0$) and $0.12$\%\ ($z=2.5$) for the smallest box size (II-A); $0.07$\%\ and $0.09$\%\ for the intermediate box size (II-B); $0.08$\%\ and $0.11$\%\ for the large box size (II-D); and $0.08$\%\ and $0.12$\%\ for the largest box size (II-F).

\subsection{Some caveats and uncertainties}
\label{ss:uncert}

It is important to remember several caveats of this analysis. First is the small box size. It is difficult to rigorously test convergence, but the small and statistically insignificant change in $b_{Fv}$ going from II-B to II-F (with $27\times$ the simulation volume) is consistent with a large box size limit being reached. Nevertheless, even Box II-F (side length 2.55 cMpc) has a total mass of $6.5\times 10^{11}\,M_\odot$, which limits the size of structures that can form and implies a missing large-scale variance of $\sigma^2(M) = 0.77$ at $z=2.5$. (The missing variance is 2.67 for Box II-A.) A related issue is that we are stitching the biasing coefficients measured in these small-box simulations together with large-scale structure perturbation theory. This is unavoidable given present limitations: a box with Phase II resolution but, say, 4 BAO scale lengths on a side would have $1.4\times 10^{15}$ particles. Thus in future work a more rigorous study of the uncertainties in the mixed approach is desirable.

A second issue is the initial conditions. {\sc Gadget 2} is designed to handle baryons and dark matter, but at the decoupling epoch, photons and neutrinos together are 24\%\ of the Universe. A fully correct set of initial conditions would include these contributions in the background expansion rate of the Universe, follow the non-instantaneous kinematic decoupling of baryons from photons, and appropriately re-adjust the initial amplitude of dark matter perturbations. It would also be desirable to use a methodology other than the fixed comoving gravitational softening length, since the large softening length required to suppress spurious dynamical friction is not optimal for following the formation of dense structures.

A third caveat is the simulation physics. A range of physics formulations were investigated in this paper, with only minor changes in the streaming velocity bias $b_{Fv}$. However, there are aspects of the physics that we did not consider. One is that small-scale structure itself increases the clumping factor and hence the number of recombinations, and thus delays reionization and makes the IGM more transparent.
\cchange{
While this is generally expected to be a small effect \citep[e.g.][]{2006MNRAS.366..689C}, this ``indirect'' effect goes in the same direction as the ``direct'' effect of streaming velocities considered in this paper. We see from the sensitivities in Table~\ref{tab:phase2} that the indirect effect would be the same as the direct effect if the extra clumpiness of the $v_{\rm bc}=0$ case delayed reionization by $-\Delta z \sim 0.22$.\footnote{\cchange{For the II-F box size, at $z=2.5$, we calculate $\Delta\ln\tau_1 = 0.00300$ for turning off the streaming velocities, and $-d\ln\tau_1/dz = [0.02223-(-0.00505)]/2 = 0.01364$; then $0.00300/0.01364 = 0.22$.}} We have made an order-of-magnitude estimate of this effect using the ``minimal reionization model'' described in Section 9.3 of \citet{2012ApJ...746..125H}. For a given clumping factor history $C(z)$, and given an ionizing source emissivity, the model uses a single ordinary differential equation (ODE) to predict the volume filling fraction of ionized gas, $Q_{\rm HII}(z)$. We extracted the clumping factor $C_{100}(z)$ in the large (II-F) neutral simulation boxes with streaming velocities and have interpolated across the snapshots; $C_{100}$ rises from 4.6 at $z=12$ to 10.1 at $z=6$. We further re-scale the ionizing source emissivity of \citet{2012ApJ...746..125H} by a factor of 1.58 so that the midpoint of reionization ($Q_{\rm HII}=0.5$) occurs at $z=8.0$. By swapping in the clumping factor {\em without} streaming velocities (II-Fn simulation), and keeping the ionizing emissivity the same, we find that the midpoint of reionization is delayed to $z=7.5$, i.e.\ a delay of $-\Delta z =0.5$. Taken at face value, this would imply that the indirect effect is 2.3 times larger than the direct effect. However, we believe this is a significant overestimate, since once the gas is ionized the clumping factor is reduced \citep[e.g.][]{2009MNRAS.394.1812P}. In our simulations, the difference $C_{100}({\rm without\,}v_{\rm bc})-C_{100}({\rm with\,}v_{\rm bc})$ decays after a region is reionized, dropping to $1/e$ of its initial value after 20 Myr. Since the duration of reionization was probably longer than 20 Myr\footnote{\cchange{Most of the parameter space for rapid reionization is now excluded by null results from 21 cm observations; see \citet{2017ApJ...847...64M}.}}, in reality we expect that the streaming velocities only affected the clumping factor in the subregions of the ionized bubbles that recently reionized. An investigation of this aspect is beyond the capabilities of the single ODE model, but we expect that it would reduce the indirect effect of streaming velocities modulating clumping and hence reionization.
Also, this} indirect effect would be non-local, however, and so might not produce a BAO peak shift that can be modeled in the formalism of \citet{2016PhRvL.116l1303B}.
We also considered only hydrodynamics, and neglected any dynamical effects from magnetic fields and cosmic rays.

We also did not consider He\,{\sc ii} reionization, which is believed to have occurred around $z\sim 3.5$, i.e.\ before the epoch of most of the BOSS observations. This can have a substantial impact on how quickly IGM gas approaches a simple temperature-density relation and forgets its initial thermal state. There are two competing effects: on the one hand, He\,{\sc ii} photoionization heating in steady state (i.e.\ inside a He\,{\sc iii}-dominated region) acts to speed up the approach to the $T-\Delta$ relation, but the added energy injection in the He\,{\sc ii} $\rightarrow$ He\,{\sc iii} transition itself acts to slow this down because the heated gas experiences fewer recombinations. As a specific but very simple example, let us consider the model of \citet{2016MNRAS.456...47M}, initialized at $T=2\times 10^4$ K at $z=8$. By varying the initial conditions, one finds that $\partial\ln T(z=2.5)/\partial \ln T(z=8) = 0.031$ with He\,{\sc ii} reionization neglected. If He\,{\sc ii} reionization is turned on at $z=3.5$, accompanied by the instantaneous injection of 43 eV of energy per He atom, one finds that $\partial\ln T(z=2.5)/\partial \ln T(z=8) = 0.017$, i.e.\ the final temperature of the gas is less sensitive to initial conditions. A full model would also take into account the non-instantaneous energy injection in He\,{\sc ii} reionization due to soft X-rays that pre-heat the singly ionized IGM before the arrival of an ionization front (see \citealt{2016MNRAS.460.1885U} for a recent example). A full exploration of the impact of He\,{\sc ii} reionization on the streaming velocity sensitivity is beyond the scope of the present work.

Turning now to the interpretation of the simulations outputs rather than the simulations themselves, we note that Eq.~(\ref{eq:stream}) is valid in real-space. However the redshift-space effects in both the Lyman-$\alpha$ forest and the streaming velocity terms are significant. The conventional 3D Lyman-$\alpha$ forest power spectrum (Eq.~\ref{eq:Linear}) is enhanced by a factor of $(1+\beta_F\mu^2)^2$ relative to real-space theory. The dominant term in the streaming velocity contamination is the advection term \citep[Eq.~A6]{2016PhRvL.116l1303B}, which -- repeating the derivation in Appendix A of \citep{2016PhRvL.116l1303B}, but replacing the advection term with the displacement from Lagrangian to {\em redshift} space -- will be enhanced by a factor of $(1+\beta_F\mu^2)(1+f\mu^2)$, where $f$ is the normalized growth rate of structure. Since $f\approx 1$ in the matter-dominated era, and coincidentally $\beta_F\approx 1$, it may be that the distortion of the BAO peak in redshift space is similar to that in real space. However, more work is required to be sure, since the BAO peak shift due to streaming velocities also includes other terms, notably those that depend on the second-order bias coefficients ($b_2$ and its redshift-space analogues).\footnote{We plan a more detailed investigation of this issue in a future paper (Givans et~al., in prep.).} To our knowledge these have not been reported for the Lyman-$\alpha$ forest in either observations or simulations.

Finally, this paper has not considered any kind of speculative feedback mechanisms by which small-scale structure could affect the reheating or reionization of the IGM -- we have assumed that most of the mini-haloes remain sterile and do nothing except get destroyed. While this is the conventional view and is well-motivated by the inefficiency of atomic cooling in these haloes, one should keep in mind that it could turn out to be incorrect.

In summary, while the results for the BAO peak shift predicted here are based on a reasonable first set of simulations, there are several conventional sources of error that could plausibly be at the factor of a few level. The 0.10\%\ shift prediction should thus be treated with some caution until these issues are addressed. This is in addition to the ``unknown unknowns'' that necessarily exist given the range of scales and redshifts involved.

\section{Discussion}
\label{sec:discussion}

This paper has made a first estimate of the Lyman-$\alpha$ forest BAO scale shift due to primordial streaming velocities. The BAO scale shift depends on how much the primordial streaming velocity changes the transmitted flux of the Lyman-$\alpha$ forest.

We explored the streaming velocity effects using a suite of hydrodynamic simulations. As found in previous studies, streaming velocities modulate the amount of small-scale structure that forms prior to reionization. The way in which this small-scale structure affects the Lyman-$\alpha$ forest is more subtle. Small scales go fully non-linear before reionization, and thus lead to a primordial cosmic web of mini-halos, mini-filaments, and mini-voids that are below the ionized-gas Jeans scale. These structures are destroyed by reionization, but are not forgotten: their impact on the thermal and dynamical structure of the IGM can persist for many Gyr. The principal impact on the thermal state of the gas is that material in mini-voids has high entropy immediately following reionization, and remains above the mean temperature-density relation even at $z<4$. This behavior can be quantitatively explained by simple analytical models \citep{2016MNRAS.456...47M}, and is a direct consequence of any structure formation theory with power at small scales and in which the IGM is reheated by ionization fronts. Ultimately, simulation boxes with higher streaming velocities have less small-scale baryonic structure at reionization, lower entropy at late times, and lower transmitted flux.

By changing the streaming velocity in small-box simulations, we can infer the sensitivity coefficients and forecast a 0.12\%\ change in the Lyman-$\alpha$ forest BAO peak position at $z=2.5$. This should be interpreted as only an order of magnitude prediction -- it remains very uncertain, as it is stitched together from a combination of small hydrodynamic simulations and large-scale structure perturbation theory, and some important aspects of the physics (e.g.\ He\,{\sc ii} reionization) and statistics (e.g.\ redshift-space distortions) are not treated in a fully consistent way. We find much larger sensitivity coefficients for the dependence of the Lyman-$\alpha$ transmission on the redshift of reionization, however reionization bubbles are not expected to couple to the BAO scale, except possibly indirectly via the streaming velocities. Such indirect pathways should be investigated in future work.

For comparison, the current Lyman-$\alpha$ forest BAO constraint from BOSS is that the standard ruler length is $100\%\times \alpha_\parallel = 105.3\pm 3.6$\%\ and $100\%\times \alpha_\perp = 96.5\pm 5.5$\%\ times the length expected based on the CMB data and the $\Lambda$CDM cosmological model in the radial and transverse directions, respectively \citep{2017arXiv170200176B}. There is also a BAO measurement from the cross-correlation of the Lyman-$\alpha$ forest with quasars in BOSS, yielding \cchange{$100\%\times \alpha_\parallel = 107.7\pm 4.2$\%\ and $100\%\times \alpha_\perp = 89.8\pm 4.2$\%\ \citep{2017arXiv170802225D}}. The shifts predicted in this paper are therefore too small to substantially affect the BOSS results, or to explain the tension of marginal statistical significance in which $\alpha_\parallel>1$ whereas $\alpha_\perp<1$.

Lyman-$\alpha$ BAO constraints will improve considerably in the near future. The planned DESI program will measure the BAO scale using the Lyman-$\alpha$ forest to an aggregate precision of 0.46\%, with the centroid of the redshift weight at $z=2.44$ \citep[Table 2.7]{2016arXiv161100036D}. The predicted peak shift is thus a $\sim 0.26\sigma$ effect for DESI, which would make it a minor but not negligible correction.

While DESI is the largest approved Lyman-$\alpha$ forest survey, it is still far from being cosmic variance limited (this limit is $\sim$0.07\%\ in the $2<z<3$ range; \citealt{2007ApJ...665...14S}, Figure 3), so if resources and technology allow it should be possible to improve the precision further. One possibility is to increase the density of lines of sight with a hyper-multiplexed spectrograph on a $\gtrsim 8\,$m telescope, likely using galaxies as backlights \citep{2014ApJ...795L..12L}, as has been suggested during the Department of Energy Cosmic Visions process \citep{2016arXiv160407626D}.\footnote{Kyle Dawson, Khee-Gan Lee, and An\v ze Slosar, private communication.} If any such ambitious project comes to fruition, then it may be essential to use higher-order statistics \citep[e.g.][]{2011JCAP...07..018Y, 2015MNRAS.448....9S} to measure and correct the streaming velocity shift in the BAO peak. \cchange{In particular, the streaming velocity effect has a very specific imprint on the angular structure of the 3-point function \citep{2015MNRAS.448....9S}.}

This paper has presented a first attempt to predict the order of magnitude of the streaming velocity bias $b_v$ and the BAO peak shift $\Delta\alpha$ in the Lyman-$\alpha$ forest. However, there are ambitious plans to measure the BAO scale using other tracers, including emission line galaxies (e.g.\ with DESI, \citealt{2016arXiv161100036D}; the Subaru Prime Focus Spectrograph, \citealt{2014PASJ...66R...1T}; 4MOST, \citealt{2016SPIE.9908E..1OD}; {\slshape Euclid}, \citealt{2011arXiv1110.3193L}; and {\slshape WFIRST}, \citealt{2015arXiv150303757S}) and H\,{\sc i} intensity maps (e.g.\ with CHIME, \citealt{2014SPIE.9145E..22B}; HIRAX, \citealt{2016SPIE.9906E..5XN}; BINGO, \citealt{2013MNRAS.434.1239B}; and more ambitious follow-on experiments). In these cases, the underlying tracers are galaxies (individually detected or not), and hence predictions for $b_v$ are complicated by star formation and feedback. Despite these added complications, the importance of these tracers motivates further study of the range of possible models, the resulting streaming velocity biases, and their implications for the BAO feature.

\section*{Acknowledgements}

C.H.\ thanks Jonathan Blazek, Joseph McEwen, Molly Peeples, Zachary Slepian, and David Weinberg for helpful discussions and feedback; and Jonathan Blazek for re-running the scripts from \citet{2016PhRvL.116l1303B}.
C.H.\ is supported by the David \& Lucile Packard Foundation, the Simons Foundation, the U.S.\ Department of Energy, and the National Aeronautics and Space Administration.
\cchange{C.H.\ thanks the anonymous referee for comments that improved the paper. The computations in this paper were run on the CCAPP condo of the Ruby Cluster at the Ohio Supercomputer Center.}

\appendix

\section{Density-dependent reionization temperature model}
\label{app:I-front}

The main text of this article uses two models for the gas temperature $T_{\rm re}$ immediately following reionization. The simplest model is to assume that the reionization temperature is independent of density, which is common in studies of the thermal history of the IGM. This appendix describes the alternative model, which is a physical model for the density dependence $T_{\rm re}(\Delta)$ determined by the speed of an ultraviolet-driven ionization front and the balance of photo-ionization and collisional cooling within the (finite) front width. It follows the physical reasoning of \citet{1994MNRAS.266..343M}.

The model is a simple 1D time-dependent ionization front. The depth parameter is the {\em total} hydrogen column $N_{\rm H}$ (units: cm$^{-2}$). A grid of $N_{\rm grid}$ cells of width $\Delta N_{\rm H}$ is built, with each cell $j\in\{0...N_{\rm grid}-1\}$ containing a hydrogen neutral fraction $y_{{\rm H1},j}$, a helium neutral fraction $y_{{\rm He1},j}$, and an energy per hydrogen nucleus $E_j$. Physical distance is related to $N_{\rm H}$ by $x=N_{\rm H}/n_{\rm H}$, where $n_{\rm H}$ is the 3D hydrogen number density (assumed constant). Incident on the left ($N_{\rm H}=0$) side of the grid is a flux of ionizing photons $F$ (units: photons cm$^{-2}$ s$^{-1}$). The theoretical velocity of the ionization front is $v_{\rm i} = F/[n_{\rm H}(1+f_{\rm He})]$, where $f_{\rm He}$ is the helium:hydrogen ratio by number. We introduce a scaled time parameter $t' = Ft$, with units of photons cm$^{-2}$. In the re-scaled coordinates, the ionization front is expected to proceed at a speed
\begin{equation}
\frac{{\rm d}N_{\rm H}}{{\rm d}t'} = \frac{{\rm d}N_{\rm H}/{\rm d}x}{{\rm d}t'/{\rm d}t} v_{\rm i} 
= \frac{1}{1+f_{\rm He}}.
\end{equation}

The incident flux is broken into a set of frequency bins $\alpha\in\{0...N_\nu-1\}$, spanning the range from 1 to 4 Ry with logarithmic spacing. Photons at $\nu<1$ Ry are non-ionizing and do not need to be tracked, while those at $\nu>4$ Ry are He{\,\sc ii}-ionizing and assumed to be blocked at a He{\,\sc ii} ionization front much closer to the source. Each bin contains a fraction $f_\alpha$ of the total {\em photon} flux. For a blackbody incident spectrum, we have
\begin{equation}
f_\alpha \propto \frac{\nu_\alpha^3}{{\rm e}^{h\nu_\alpha/k_{\rm B}T_{\rm bb}} - 1},
\end{equation}
with the proportionality given by the normalization condition $\sum_{\alpha=0}^{N_\nu-1} f_\alpha = 1$. This flux is attenuated by photo-ionization: each cell $j$ provides an optical depth to photons in frequency bin $\alpha$ given by $\tau_{j\alpha} = \tau_{j\alpha}^{\rm HI} + \tau_{j\alpha}^{\rm HeI}$, with
\begin{equation}
\tau_{j\alpha}^{\rm HI} = \Delta N_{\rm H}\, y_{{\rm H1},j} \sigma_\alpha^{\rm HI}
~~{\rm and}~~
\tau_{j\alpha}^{\rm HeI} = f_{\rm He} \Delta N_{\rm H} \, y_{{\rm He1},j} \sigma_\alpha^{\rm HeI}.
\end{equation}
The H{\,\sc i} cross section is calculated from the exact nonrelativistic dipole result, while for He{\,\sc i} we use the fitting function of \citet{1996ApJ...465..487V}. The He{\,\sc i} cross section and hence $\tau_{j\alpha}^{\rm HeI}$ are zero below the He{\,\sc i} threshold at $h\nu<24.6$ eV.

Within each cell $j$ and each frequency bin $\alpha$, there is a number of absorbed photons per hydrogen nucleus per rescaled time (i.e.\ per d$t'$) given by
\begin{equation}
{\cal A}_{j\alpha} = f_\alpha \exp \Bigl(-\sum_{j'=0}^{j-1} \tau_{j'\alpha} \Bigr) \frac{ 1-\exp(-\tau_{j\alpha}) }{\Delta N_{\rm H}}.
\label{eq:calA}
\end{equation}
This results in photo-ionization rates
\begin{equation}
\frac{{\rm d}y_{{\rm H1},j}}{{\rm d}t'} = \sum_{\alpha=0}^{N_\nu-1} {\cal A}_{j\alpha} \frac{\tau_{j\alpha}^{\rm HI}}{\tau_{j\alpha}}
~~~{\rm and}~~~
\frac{{\rm d}y_{{\rm He1},j}}{{\rm d}t'} = \sum_{\alpha=0}^{N_\nu-1} \frac1{f_{\rm He}} {\cal A}_{j\alpha} \frac{\tau_{j\alpha}^{\rm HeI}}{\tau_{j\alpha}}.
\label{eq:dy}
\end{equation}
Note that for small $\tau_{j\alpha}$, these equations can have a 0/0 indeterminate form; in such cases, the $1-\exp(-\tau_{j\alpha})$ in Eq.~(\ref{eq:calA}) must be pulled into Eq.~(\ref{eq:dy}), and the expansion $(1-{\rm e}^{-\tau})/\tau \rightarrow 1 - \frac12\tau + ...$ is used.

Next we consider the heating and cooling of the gas in the ionization front. The temperature is given by
\begin{equation}
T_j = \frac{2E}{3k_{\rm B}[2-y_{{\rm H1},j} + f_{\rm He}(2-y_{{\rm He},j}) ]}
\end{equation}
and the electron abundance per hydrogen nucleus is
\begin{equation}
x_{e,j} = 1-y_{{\rm H1},j} + f_{\rm He}(1-y_{{\rm He1},j}).
\end{equation}
The net heating rate is
\begin{eqnarray}
 \frac{{\rm d}E_j}{{\rm d}t'} 
\!\! &=& \!\! \sum_{\alpha=0}^{N_\nu-1} {\cal A}_{j\alpha} \frac{\tau_{j\alpha}^{\rm HI}(h\nu-I_{\rm HI})
+ \tau_{j\alpha}^{\rm HeI}(h\nu-I_{\rm HeI})
}{\tau_{j\alpha}}
\nonumber \\
&& \!\! - \frac{ y_{{\rm H1},j}x_{e,j} }{v_{\rm i}(1+f_{\rm He})}  \sum_{n=2}^3 q_{1\rightarrow n} h\nu_{\rm I} (1-n^{-2}),
\end{eqnarray}
where $I_{\rm HI}$ and $I_{\rm HeI}$ denote the ionization energies. The first term here describes photo-ionization, and the second collisional cooling due to excitation of H{\,\sc i}; the amount of energy lost by exciting a hydrogen atom to the $n$th level is $h\nu_{\rm I}(1-n^{-2})$ according to the Rydberg formula. The collisional excitation rate coefficients (units: cm$^3$ s$^{-1}$) are taken from the fitting formulae of \citet{1983MNRAS.202P..15A}. The factor of $v_{\rm i}(1+f_{\rm He}) = F/n_{\rm H}$ arises from conversion of $t$ to rescaled time $t'$, and from the conversion of absolute particle densities to particles per hydrogen nucleus.

The above calculation results in a system of $3N_{\rm grid}$ ordinary differential equations. Due to the simplicity of this system, even the first-order Euler method is fast enough; convergence to 4 decimal places is typically obtained with steps of $\Delta t' = 2.5\times 10^{15}$ cm$^{-2}$ and $\Delta N_{\rm H} = 2.5\times 10^{16}$ cm$^{-2}$. We used $N_\nu=128$ frequency bins. The system is initialized to be cold and neutral, and is evolved until $t'=3\times 10^{19}$ cm$^{-2}$ -- long enough to build a steady-state ionization front that marches across the grid, and cleanly separate the ionization front itself from the initial thermal structure left behind at $N_{\rm HI}\lesssim 5\times 10^{18}$ cm$^{-2}$.

The above calculation neglects losses due to secondary excitations and ionizations, which would be a poor approximation for X-rays but is valid in the ultraviolet range. Even for a 28 eV primary photo-electron (i.e.\ incident photon energy of $13.6+28=41.6$ eV), \citet{1985ApJ...298..268S} estimate such losses to be 20 per cent at $x_e=0.035$, declining to 10 per cent at $x_e=0.09$. The probability of a secondary ionization of hydrogen is estimated to be 5 per cent even at $x_e=0.022$.

For an incident spectrum proportional to a $5\times 10^4$ K blackbody, and an ionization front velocity of $v_{\rm i} = 10^2,\ 10^3,\ 10^4$ km s$^{-1}$ (0.1, 1, 10 cMpc per $\Delta z=1$ at $z=8$), we find a post-ionization front temperature of $T_4=1.38,\ 1.81,\ 2.17$. \cchange{The main text uses a model with a $5\times 10^4$ K blackbody incident spectrum and a flux such that the ionization front velocity is $5\times 10^8$ cm s$^{-1}$ at mean density (the velocity scales inversely with density at constant flux). The post-reionization temperature for this model is shown in Figure~\ref{fig:tdfig}.}

\begin{figure}
\includegraphics[width=3.2in]{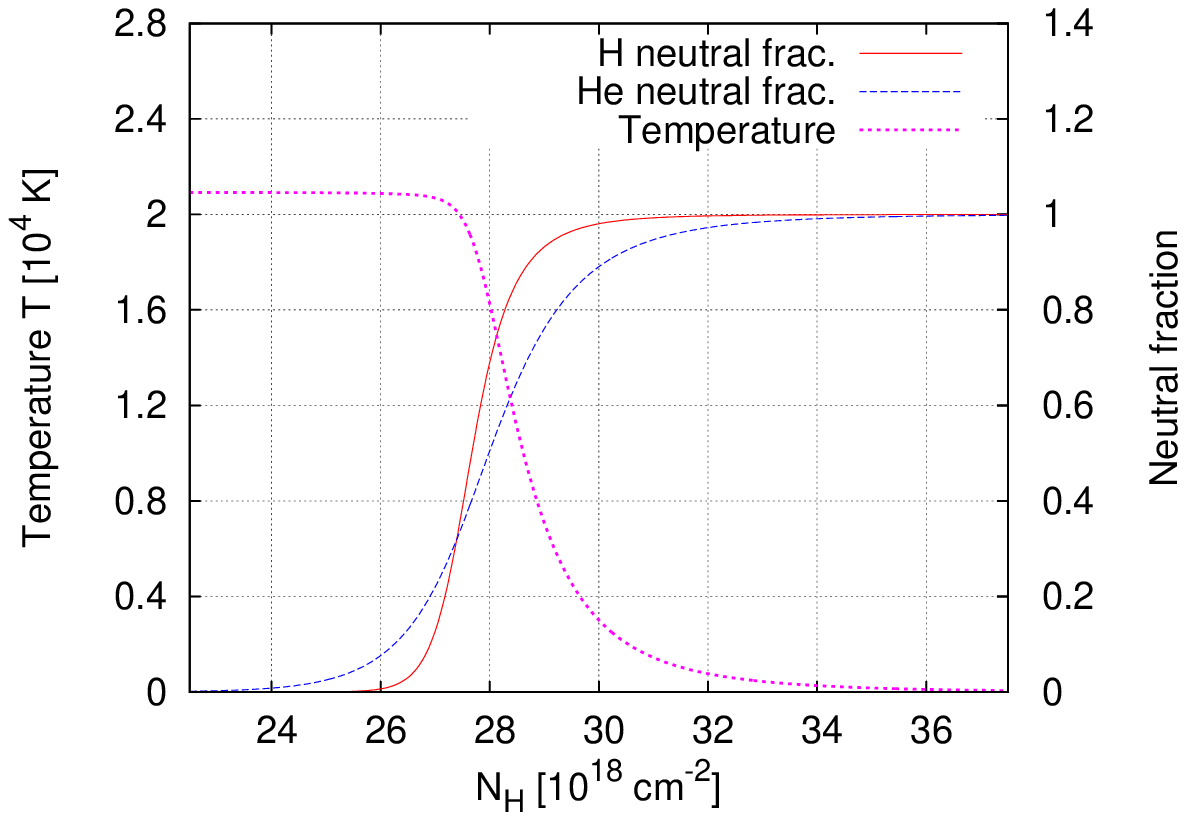}
\caption{\label{fig:ionmodel}An example ionization front model, with an incident blackbody radiation field at $5\times 10^4$ K and a front velocity of $v_{\rm i} = 4.9\times 10^3$ km s$^{-1}$.}
\end{figure}

\begin{figure}
\includegraphics[width=3.2in]{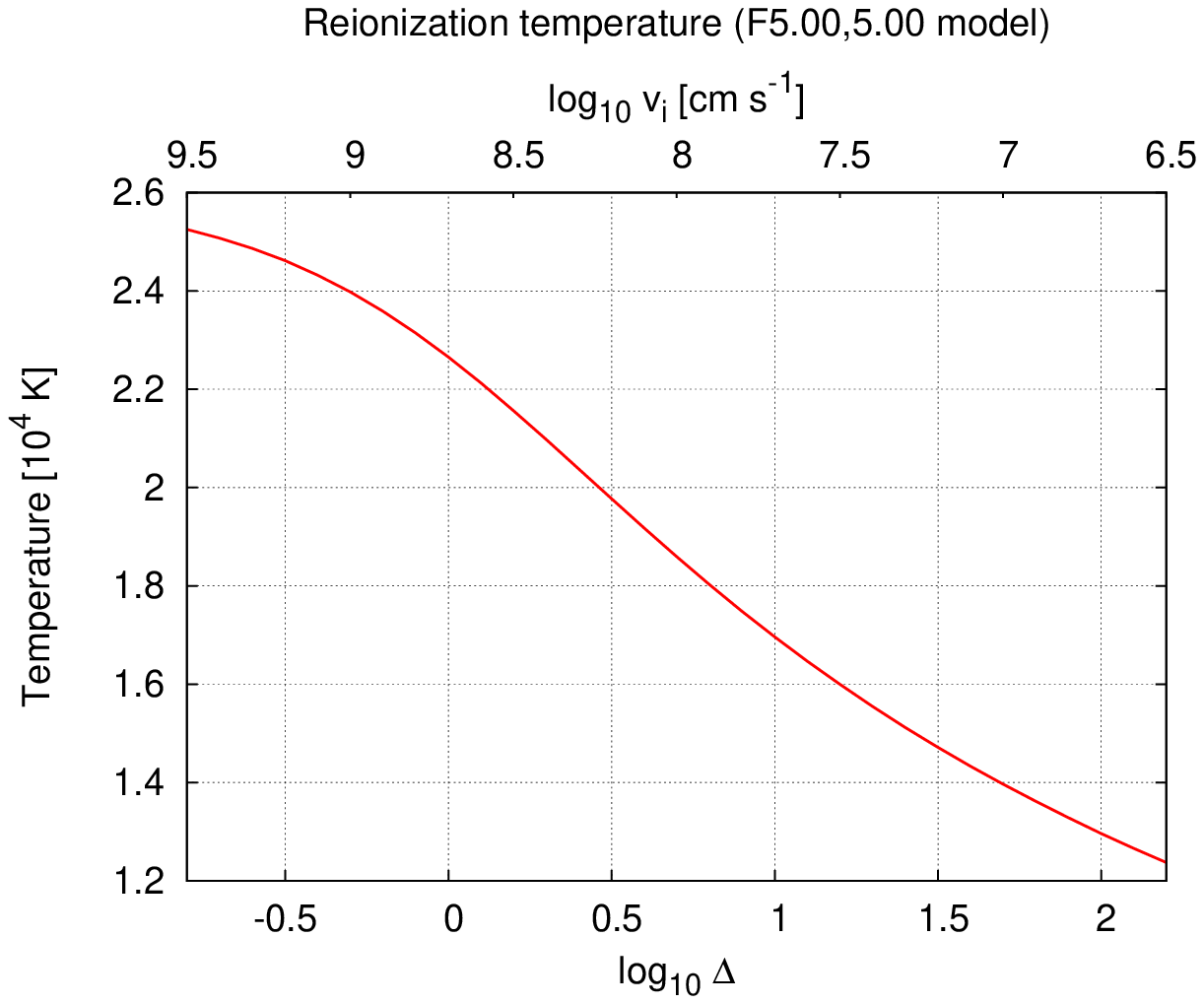}
\caption{\label{fig:tdfig}\cchange{The temperature-density relation immediately after reionization for an incident $5\times 10^4$ K blackbody spectrum and an ionization front velocity $v_{\rm i} = 5\times 10^8\Delta^{-1}$ cm s$^{-1}$ (the ``{\tt F} 5.00,5.00'' model used in the main text).}}
\end{figure}

\end{document}